\documentclass{article}

\usepackage{arxiv}

\usepackage[utf8]{inputenc} 
\usepackage[T1]{fontenc}    
\usepackage{hyperref}       
\usepackage{url}            
\usepackage{booktabs}       
\usepackage{amsfonts}       
\usepackage{nicefrac}       
\usepackage{microtype}      
\usepackage{lipsum}
\usepackage{graphicx}
\usepackage{amssymb}
\usepackage{amsmath,amsthm}
\usepackage{array}
\usepackage{subfig}
\usepackage{capt-of}
\usepackage{mathtools}
\usepackage[numbers, square, comma, sort&compress]{natbib}

\title{The effects of imitation dynamics on vaccination behaviours
	in SIR-network model}

\author{
  Sheryl L. ~Chang \textsuperscript{a}
  \thanks{Corresponding author} \\
  \texttt{ sheryl.chang@sydney.edu.au} \\
   \And
Mahendra  Piraveenan \textsuperscript{a,b}\\
  \texttt{mahendrarajah.piraveenan@sydney.edu.au} \\  
   \And
Mikhail Prokopenko \textsuperscript{a,c} \\
\vspace*{1cm}
\texttt{mikhail.prokopenko@sydney.edu.au} \\
  \textsuperscript{a}Complex Systems Research Group, Faculty of Engineering and IT \\
  The University of Sydney, NSW, 2006, Australia \\
     \textsuperscript{b}Charles Perkins Centre, The University of Sydney \\
  John Hopkins Drive, Camperdown, NSW, 2006, Australia\\
  \textsuperscript{c}Marie Bashir Institute for Infectious Diseases and Biosecurity \\
   The University of Sydney, Westmead, NSW, 2145, Australia\\
}
\begin{document}
\maketitle
\begin{abstract}
We present a series of SIR-network models, extended with a game-theoretic treatment of imitation dynamics which result from regular population mobility across residential and work areas and the ensuing interactions.  Each considered SIR-network model captures a class of vaccination behaviours influenced by epidemic characteristics, interaction topology, and imitation dynamics. Our focus is the eventual vaccination coverage, produced under voluntary vaccination schemes, in response to these varying factors.
Using the next generation matrix method, we analytically derive and compare expressions for the basic reproduction number $R_0$ for the proposed SIR-network models.
Furthermore, we simulate the epidemic dynamics over time for the considered models, and show that if individuals are sufficiently responsive towards the changes in the disease prevalence, then the more expansive travelling patterns encourage convergence to the endemic, mixed  equilibria. On the contrary, if individuals are insensitive to changes in the disease prevalence,  we find that they tend to remain unvaccinated in all the studied models. Our results concur with earlier studies in showing that residents from highly connected residential areas are more likely to get vaccinated. We also show that the existence of the individuals  committed to receiving vaccination reduces $R_0$ and delays the disease prevalence, and thus is essential to containing epidemics.
\end{abstract}


\section{Introduction}
\label{intro}
Vaccination has long been established as a powerful tool in managing and controlling infectious diseases by providing protection to susceptible individuals \cite{WHO}. With a sufficiently high vaccination coverage, the probability of the remaining unvaccinated individuals getting infected  reduces significantly. However, such systematic programs by necessity may limit the freedom of choice of individuals. When vaccination programs are made voluntary, the vaccination uptake declines as a result of individuals choosing not to vaccinate, as seen in Britain in 2003 when the vaccination program for Measles-Mumps-Rubella (MMR) was made voluntary \cite{measle_outbreak}. Parents feared possible complications from vaccination \cite{measle_outbreak,bauch_2005_imi} and hoped to exploit the `\textit{herd immunity}' by assuming other parents would choose to vaccinate their children. Such hopes did not materialize precisely because other parents also thought similarly.  

Under a voluntary vaccination policy an individual's decision depends on several factors: the social influence from one's social network, the risk perception of vaccination, and the risk perception of infection, in terms of both likelihood and impact \cite{vac_review}. This decision-making is often modelled using game theory \cite{bauch_2005_imi,imi_bauch_bha,sto_network_2010,sto_network_2012,sto_network_2012_memory,sto_network_2013,net_imi_nei,comp_imi_nei}, by allowing individuals to compare the cost of vaccination and the potential cost of non-vaccination (in terms of the likelihood and impact of infection), and adopting imitation dynamics in modelling the influence of social interactions. However, many of the earlier studies \cite{static_bauch,attractor_bauch,static_group_ind} were  based on the key assumption of a well-mixed homogeneous population where each individual is assumed to have an equal chance of making contact with any other individual in the population. This population assumption is rather unrealistic as large populations are often diverse with varying levels of interactions. To address this, more recent studies \cite{sto_network_2012,sto_network_2012_memory,sto_network_2013,net_imi_nei,comp_imi_nei,network_game_bauch} model  populations as complex networks where each individual, represented by a node, has a finite set of contacts, represented by links \cite{vac_net_review}.  

It has been shown that there is a critical cost threshold in the vaccination cost above which the likelihood of vaccination  drops steeply \cite{sto_network_2010}. In addition, highly connected individuals were shown to be more likely to choose to be vaccinated as they perceive themselves to be at a greater risk of being infected due to high exposure within the community. In modelling large-scale epidemics, the population size (i.e., number of nodes) can easily reach millions of individuals, interacting in a complex way. The challenge,  therefore, is to extend  epidemic modelling not only with the individual vaccination decision-making, but also capture diverse interaction patterns encoded within a network. Such an integration has not yet been formalized, motivating our study. Furthermore, once an integrated model is developed, a specific challenge is to consider how the vaccination imitation dynamics developing across a network affects  the basic reproduction number, $R_0$. This question forms our second main objective.

One relevant approach partially addressing this objective is offered by the multi-suburb (or multi-city) SIR-network model \cite{ArinoR0,multi-city,patch} where each node represents a neighbourhood (or city) with a certain number of residents. 
The daily commute of individuals between two neighbourhoods is modelled along the network link connecting the two nodes, which allows to quantify the disease spread between individuals from different neighbourhoods. Ultimately, this model captures meta-population dynamics in a multi-suburb setting affected by an epidemic spread at a greater scale. To date, these models have not yet considered intervention (e.g., vaccination) options, and the corresponding social interaction across the populations.

To incorporate the imitation dynamics, modelled game-theoretically, within a multi-suburb model representing mobility, we propose a series of integrated vaccination-focused SIR-network models. This allows us to systematically analyze  how travelling patterns affect the voluntary vaccination uptake due to adoption of different imitation choices, in a large distributed  population. The developed models  use an increasingly complex set of  vaccination strategies. Thus, our specific contribution  is the study of the vaccination uptake, driven by imitation dynamics under a voluntary vaccination scheme, using an SIR model on a complex network representing a multi-suburb  environment, within which the individuals commute between residential and work areas.

In section \ref{method}, we present the models and methods associated with this study: in particular, we present  analytical derivations of the basic reproduction number $R_0$  for the proposed models using the Next Generation Operator Approach \cite{NGM1990}, and carry out a comparative analysis of $R_0$ across these  models. In section \ref{results}, we simulate the epidemic and vaccination dynamics over time using the proposed models in different network settings, including a pilot case of a 3-node network, and a 3000-node Erd\"os-R\'enyi random network. The comparison of the produced results across different models and settings is carried out for the larger network, with the focus on the emergent attractor dynamics, in terms of the proportion of vaccinated individuals. Particularly, we analyze the sensitivity of the individual strategies (whether to vaccinate or not) to the levels of disease prevalence produced by the different considered models. Section \ref{conclusion} concludes the study with a brief discussion of the importance of these results. \\

\section{Technical background}
\subsection{Basic reproduction number $R_0$}
The basic reproduction number $R_0$ is defined as the number of secondary infections produced by an infected individual in an otherwise completely susceptible population \cite{R0_ori}. It is well-known that $R_0$ is an epidemic threshold, with the disease dying out as $R_0<1$, or becoming endemic as $R_0>1$ \cite{ArinoR0, harding2018}. This finding strictly holds only in deterministic models with infinite population \cite{Hartfield2013}. The topology of the underlying contact network is known to affect the epidemic threshold \cite{Pastor-Satorras}.  Many disease transmission models have shown important correlations between $R_0$ and the key epidemic characteristics (e.g., disease prevalence, attack rates, etc.) \cite{R0_book}.  In addition,  $R_0$ has been considered as a critical threshold for phase transitions studied with methods of statistical physics or information theory \cite{erten2017}. 

\subsection{Vaccination model with imitation dynamics}
\label{vaxx-SIR}
 Imitation dynamics, a process by which individuals copy the strategy of other individuals, is widely used to model vaccinating behaviours incorporated with SIR models. The model proposed in \cite{bauch_2005_imi} applied game theory to represent parents' decision-making about whether to get their newborns vaccinated against childhood disease (e.g., measles, mumps, rubella, pertussis). In this model, individuals are in a homogeneously mixing population, and susceptible individuals have two `pure strategies' regarding vaccination:  to vaccinate or not to vaccinate. The non-vaccination decision can change to the vaccination decision at a particular sampling rate, however the vaccination decision cannot be  changed to non-vaccination.  Individuals adopt  one of these strategies by weighing up their perceived payoffs,  measured by the probability of morbidity from vaccination, and the risk of infection respectively. The payoff for vaccination ($f_v$) is given as,
\begin{equation}
f_v=-r_v
\label{vaccinator}
\end{equation}
and the payoff for non-vaccination ($f_{nv}$), measured as the risk of infection, is given as
\begin{equation}
f_{nv}=-r_{nv}mI(t) 
\label{infection}
\end{equation}
where $r_v$ is the perceived risk of morbidity from vaccination, $r_{nv}$ is the perceived risk of morbidity from non-vaccination (i.e., infection), $I(t)$ is the current disease prevalence in population fraction at time $t$, and $m$ is the sensitivity to disease prevalence \cite{bauch_2005_imi}. 

From Equation \ref{vaccinator} and \ref{infection}, it can be seen that the payoff for vaccinated individuals is a simple constant, and the payoff for unvaccinated individuals is proportional to the extent of epidemic prevalence (i.e., the severity of the disease). 

It is assumed that life-long immunity is granted with effective vaccination. If one individual decides to vaccinate, s/he cannot revert back to the unvaccinated status. To make an individual switch to a vaccinating strategy, the payoff gain, $\Delta E=f_v-f_{nv}$ must be positive ($\Delta E>0$). Let $x$ denote the relative proportion of vaccinated individuals, and assume that an unvaccinated individual can sample a strategy from a vaccinated individual at certain rate $\sigma$, and can switch to the `vaccinate strategy' with probability $\rho \Delta E$. Then the time evolution of $x$  can be defined as:
\begin{equation}
\begin{aligned}
\dot{x} & =(1-x)\sigma x \rho \Delta E \\
        & =\delta x(1-x)[-r_v+r_{nv}mI] \\
\end{aligned}
\label{x_bauch}
\end{equation}
where $\delta=\sigma \rho$ can be interpreted as the combined imitation rate that individuals use to sample and imitate strategies of other individuals. Equation \ref{x_bauch} can be rewritten so that $x$ only depends on two parameters: 
\begin{equation}
\dot{x} =\kappa x(1-x)(-1+\omega I)\\
\label{x_bauch_final}
\end{equation}
 where $\kappa=\delta r_v$ and $\omega=mr_{nv}/r_v$, assuming that the risk of morbidity from vaccination, $r_v$, and the risk of morbidity from non-vaccination (i.e., morbidity from infection), $r_{nv}$, are constant during the course of epidemics. Hence, $\kappa$ is the adjusted imitation rate, and $\omega$ measures individual's responsiveness towards the changes in disease prevalence.

The disease prevalence $I$ is determined from a simple SIR model \cite{R0_ori} with birth and death that divides the population into three health compartments: $S$ (susceptible), $I$ (infected) and $R$ (recovered). If a susceptible individual encounters an infected individual, s/he contracts the infection with transmission rate $\beta$, and thus progresses to the infected compartment, while an infected individual recovers at rate $\gamma$. It is assumed that the birth and death rate are equal, denoted by $\mu$ \cite{math_epi}. Individuals in all compartments die at an equal rate, and all newborns are added to the susceptible compartment unless and until they are vaccinated. Vaccinated newborns are moved to the recovered class directly, with a rate $\mu x$. Therefore, the dynamics are modelled as follows \cite{bauch_2005_imi}: 
\begin{equation}
\begin{aligned}
\dot{S} &=\mu(1-x)- \beta S I-\mu S \\
\dot{I} &=\beta S I- \gamma I -\mu I \\
\dot{R} &=\mu x+\gamma I-\mu R \\
\dot{x} &=\kappa x(1-x)(-1+\omega I)\\
\end{aligned}
\label{bauch_ori}
\end{equation}

where $S$, $I$, and $R$ represent the proportion of susceptible, infected, and recovered individuals respectively (that is, $S+I+R=1$), and $x$ represents the proportion of vaccinated individuals within the susceptible class at a given time. The variables $S$, $I$, and $R$, as well as $x$, are time-dependent, but for simplicity,  we  omit  subscript $t$ for these and other time-dependent variables.

Model \ref{bauch_ori} predicts oscillations in vaccine uptake in response to changes in disease prevalence. It is found that oscillations are more likely to occur when individuals imitate each other more quickly (i.e., higher $\kappa$).  The oscillations are observed to be more volatile when people alter their vaccinating behaviours promptly in response to changes in disease prevalence (i.e., higher $\omega$). Overall, higher $\kappa$ or $\omega$ produce stable limit cycles at greater amplitude. Conversely, when individuals are insensitive to changes in disease prevalence (i.e., low $\omega$), or imitate at a slower rate (i.e., low $\kappa$), the resultant vaccinating dynamics converge to equilibrium \cite{bauch_2005_imi}.

\subsection{Multi-city epidemic model}
\label{multicity}
\begin{figure}
	\centering
	\includegraphics[width=0.5\columnwidth]{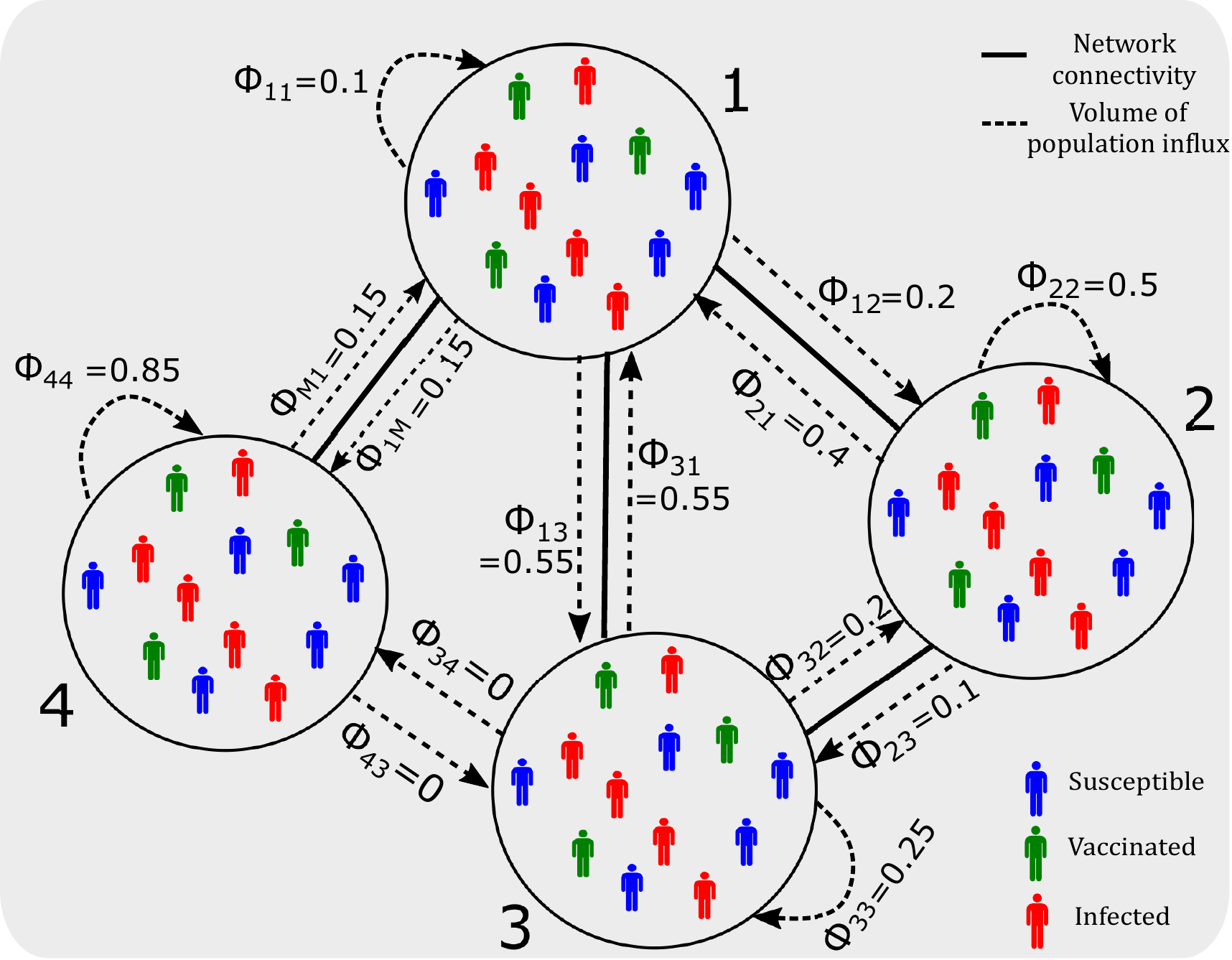}
	\caption{Schematic of daily population travel dynamics across different suburbs (nodes): a 4-node example. Solid line: network connectivity. Dashed line: volume of population flux (influx and outflux). Non-connected nodes have zero population flux (e.g., $\phi_{34}=\phi_{43}=0$). For each node, the daily outflux proportions (including travel to the considered node itself) sum up to unity, however, the daily influx proportions do not.}
	\label{influx}
\end{figure}

Several multi-city epidemic models \cite{ArinoR0, multi-city, patch} consider a network of suburbs as $M$ nodes, in which each node $i \in V$ represents a suburb,  where  $i={1,2,\cdots M}$.  If two nodes $i,j \in V$ are linked, a fraction of population living in node $i$ can travel to node $j$ and back (commute from $i$ to $j$, for example for work) on a daily basis. The connectivity of suburbs and the fraction of people commuting between them are represented by the population flux matrix $\phi$ whose entries represent the fraction of population daily commuting from $i$ to $j$, $\phi_{ij} \in [0,1]$  (Equation \ref{phi}). Note that  $\phi_{ij} \neq \phi_{ji}$, thus $\phi$ is not a symmetric matrix. Figure \ref{influx} shows an example of such travel dynamics.\\

\begin{equation}
\phi_{M \times M} =
\begin{bmatrix}
\phi_{11} & \phi_{12} & \cdots & \phi_{1M} \\
\phi_{21} & \phi_{22} & \cdots & \phi_{2M} \\
\vdots    & \vdots    & \ddots & \vdots \\
\phi_{M1} & \phi_{N2} & \cdots & \phi_{MM} 
\end{bmatrix}
\label{phi}
\end{equation}

Trivially, each row in $\phi$, representing proportions of the population of a suburb commuting to various destinations, has to sum up to unity:
\begin{equation}
\sum_{j=1}^{M}\phi_{ij}=1 \ , \ \ \forall i \in V
\end{equation}
However, the column sum measures the population influx to a suburb during a day, and therefore depends on the node's connectivity. Column summation in $\phi$ does not necessarily equal to unity. 

It is also necessary to differentiate between \textit{present population} $N_j^p$ and \textit{native population} $N_j$ in node $j$ on a particular day. $N^p_j$ is used as a normalizing factor to account for the differences in population flux for different nodes:
\begin{equation}
N^p_j =\sum_{k=1}^{M} \phi_{kj}N_k
\label{N_p}
\end{equation}

Assuming the disease transmission parameters are identical in all cities, the standard incidence can be expressed as \cite{ArinoR0}:
\begin{equation}
\sum\limits_{j=1}^M \sum\limits_{k=1}^{M} \lambda_j \beta_j \frac{I_k}{N^p_j} S_i
\label{patch_I}
\end{equation}
where $I_l$ and $S_l$ denote respectively  the number of infective and susceptible individuals in city $l$, and $\lambda_l$ is the average number of contacts in city $l$ per unit time.
Incorporating travelling pattern defined by $\phi$, yields \cite{patch}:
\begin{equation}
\sum\limits_{j=1}^M \sum\limits_{k=1}^{M} \beta_j \phi_{ij} \frac{\phi_{kj}I_k}{N^p_j} S_i
\label{patch_I_phi}
\end{equation}

The double summation term in this expression captures the infection at suburb $j$ due to the encounters between the residents from suburb $i$ and the residents from suburb $k$ ($k$ could be any suburb including $i$ or $j$) occurring at suburb $j$, provided that suburbs $i,j,k$ are connected with non-zero population flux entries in $\phi$.

$R_0$ can be derived using the Next Generation Approach (see Appendix \ref{app1} for more details). For example, for a multi-city SEIRS model with 4 compartments (susceptible, exposed, infective, recovered), introduced in \cite{ArinoR0}, and a special case where the contact rate $\lambda_j$ is set to 1, while $\beta$ is identical across all cities, $R_0$ has the following analytical solution:
\begin{equation}
R_0=\frac{\beta \epsilon}{(\gamma+d)(\epsilon+d)}
\label{Adino_R0}
\end{equation}
where $1/d,1/\epsilon,$ and $1/\gamma$ denote the average lifetime, exposed period and infective period.
It can be seen that this solution concurs with a classical SEIRS model with no mobility. When $\epsilon \to \infty$, Equation \ref{Adino_R0} reduces to $\beta / (d+\gamma)$, being a solution for a canonical SIR model. Furthermore, if the population dynamics is not considered (i.e., $d=0$), Equation \ref{Adino_R0} can be further reduced to $\beta /\gamma$, agreeing with \cite{patch}.

\section{Methods}
\label{method}
\subsection{Integrated model}
Expanding on models described in sections \ref{vaxx-SIR} and \ref{multicity}, we bring together population mobility and vaccinating behaviours in a network setting, and propose three  extensions within an integrated vaccination-focused SIR-network model: 
\begin{itemize}
	\item vaccination is available to newborns only (model \ref{system2.1_old} below )
	\item vaccination is available to the entire susceptible class (model \ref{system2.2} below)
	\item committed vaccine recipients (model \ref{system_reform} below)
\end{itemize}

In this study, we only consider vaccinations that confer lifelong immunity, mostly related to childhood diseases such as measles, mumps, rubella and pertussis. Vaccinations against such diseases are often administered for individuals at a young age, implying that newborns (more precisely, their parents) often face the vaccination decision. These are captured by the first extension above.  However, during an outbreak, adults may  face the vaccination decision themselves if they were not previously vaccinated. For example, diseases which are not included in formal childhood vaccination programs, such as smallpox \cite{smallpox}, may present vaccination decisions to the entire susceptible population. These are captured by the second extension. The third extension introduces a small fraction of susceptible population as committed vaccine recipients, i.e., those who would always choose to vaccinate regardless. The purpose of these committed vaccine recipients is to demonstrate how successful immunization education campaigns could affect vaccination dynamics.   

Our models divide the population into many homogeneous groups \cite{R0_beta}, based on their residential suburbs. Within each suburb (i.e., node), residents are treated as a homogeneous population. It is assumed that the total population within each node is conserved over time. The dynamics of epidemic and vaccination at each node, are referred to as `local dynamics'. The aggregate epidemic dynamics of the entire network can  be obtained by summing over all nodes, producing `global dynamics'.

We model the `imitation dynamics'  based on the individual's travelling pattern and the connectivity of their node defined by Equation \ref{phi}.
For any node $i \in V$, let $x_i$ denote the fraction of vaccinated individuals in the susceptible class in suburb $i$. On a particular day, unvaccinated susceptible individuals $(1-x_i)$ commute to suburb $j$ and encounter vaccinated individuals from node $k$ (where $k$ may be $i$ itself, $j$ or any other nodes) and imitate their strategy.  However, this `imitation' is only applicable in the case of a non-vaccinated individual imitating the strategy of a vaccinated individual (that is, deciding to vaccinate), since the opposite `imitation' cannot occur. Therefore, in our model, every time a non-vaccinated person from $i$ comes in contact with a vaccinated person from $k$, they imitate the `vaccinate' strategy if the perceived payoff outweighs the non-vaccination strategy.

Following Equation \ref{x_bauch_final} proposed in \cite{bauch_2005_imi}, the rate of change of the proportion of vaccinated individuals in $i$, that is, $x_i$, over time can  be expressed by:
\begin{equation}
\begin{aligned}
\dot{x_i} &=\sigma (1-x_i) \sum_{j=1}^{M} \sum_{k=1}^{M} \phi_{ij} \rho \Delta E \phi_{kj} x_k   \\
&=\delta (1-x_i) \sum_{j=1}^{M} \sum_{k=1}^{M} \phi_{ij} (-r_v+r_{nv} m I_j) \phi_{kj} x_k \\
&=\kappa (1-x_i)\sum\limits_{j=1}^M \phi_{ij}(-1+\omega I_j)\sum\limits_{k=1}^M\phi_{kj}x_k \\
\end{aligned}
\label{imitation}
\end{equation}
The players of the vaccination game are the parents, deciding whether or not to vaccinate their children using the information of the disease prevalence collected from their daily commute.  For example, a susceptible individual residing at node $i$ and working at node $j$  uses the local disease prevalence at node $j$ to decide whether to vaccinate or not. If such a susceptible individual is infected, that individual will be counted towards the local epidemic prevalence at node $i$. 

We measure each health compartment as a proportion of the population. Hence, we define a ratio, $\epsilon_l^p$, as the ratio between  present population $N_l^p$ and the `native' population $N_l$  in node $l$ on a particular day as:
\begin{equation}
\epsilon_l^p =\frac{N_l^p}{N_l} =\frac{\sum_{q=1}^{M} \phi_{ql}N_q}{N_l}
\label{normalisation}
\end{equation}

Our main focus is to investigate the effects of vaccinating behaviours on the global epidemic dynamics. To do so, we vary three parameters:
\begin{itemize}
	\item Individual's responsiveness to changes in disease prevalence, $\omega$ 
	\item Adjusted imitation rate, $\kappa$
	\item Vaccination failure rate, $\zeta$
\end{itemize}
 While $\kappa$ and $\omega$ have had been considered in \cite{bauch_2005_imi}, we introduce a new parameter $\zeta$ as the vaccination failure rate, $\zeta \in [0,1]$ to consider the cases where the vaccination may not be fully effective. 
 The imitation component  (Equation \ref{imitation}) is common in all model extensions. While the epidemic compartments vary depending on the specific extension, Equation \ref{imitation} is used consistently to model the relative rate of change in vaccination behaviours.

\subsection{Vaccination available to newborns only}
\label{newborn}
Model (\ref{system2.1_old})  captures the scenario when vaccination opportunities are provided to newborns only:
\begin{equation}
\begin{aligned}
\dot{S_i} &=\mu[\zeta x_i+(1-x_i)]-\sum\limits_{j=1}^M \sum\limits_{k=1}^{M} \beta_j \phi_{ij} \frac{\phi_{kj}I_k}{\epsilon ^p_j} S_i-\mu S_i \\
\dot{I_i} &=\sum\limits_{j=1}^M \sum\limits_{k=1}^{M} \beta_j \phi_{ij} \frac{\phi_{kj}I_k}{\epsilon ^p_j}S_i-\gamma I_i -\mu I_i \\
\dot{R_i} &=\mu(1-\zeta) x_i+\gamma I_i-\mu R_i \\
\dot{x_i} &=\kappa (1-x_i)\sum\limits_{j=1}^M \phi_{ij}(-1+\omega I_j)\sum\limits_{k=1}^M\phi_{kj}x_k \\
\end{aligned}
\label{system2.1_old}
\end{equation}  
Unvaccinated newborns and newborns with unsuccessful vaccination $\mu[\zeta x_i+(1-x_i)]$ stay in the susceptible class. Successfully vaccinated newborns, on the other hand, move to the recovered class $\mu(1-\zeta) x_i$. Other population dynamics across health compartments follow the  model (\ref{bauch_ori}). 

Since $S+I+R=1  (\dot{S}+\dot{I}+\dot{R}=0)$, model (\ref{system2.1_old}) can be reduced to:
\begin{equation}
\begin{aligned}
\dot{S_i} &=\mu[\zeta x_i+(1-x_i)]-\sum\limits_{j=1}^{M} \sum\limits_{k=1}^{M} \beta_j \phi_{ij} \frac{\phi_{kj}I_k}{\epsilon ^p_j} S_i-\mu S_i \\
\dot{I_i} &=\sum\limits_{j=1}^M \sum\limits_{k=1}^{M} \beta_j \phi_{ij} \frac{\phi_{kj}I_k}{\epsilon ^p_j}S_i-\gamma I_i -\mu I_i \\
\dot{x_i} &=\kappa (1-x_i)\sum\limits_{j=1}^M \phi_{ij}(-1+\omega I_j)\sum\limits_{k=1}^M\phi_{kj}x_k \\
\end{aligned}
\label{system2.1}
\end{equation}

Model (\ref{system2.1}) has a disease-free equilibrium (disease-free initial condition) $(S_i^0, I_i^0)$ for which 
\begin{equation}
\begin{aligned} 
S_i^0 & =\zeta x_i^0+(1-x_i^0) \\
I_i^0 & =0 \\
\end{aligned}
\label{eq_SIR2.2}
\end{equation}
\textbf{Proposition 1.}   At the disease free equilibrium, $x^0$ has two solutions ($x^0=0$ or $x^0=1$) if $x^0$ and $S^0$ are uniform across all nodes.

\textit{Proof:}
At the disease free equilibrium, assuming $S^0_i, x_i^0$ are uniform across all nodes, and therefore: 
\begin{equation}
S^0_i =  S^0, \ x_i^0 = x^0 \quad \forall i
\label{assump1}
\end{equation}
Hence, under disease-free condition where $I^0=0$ and $\dot{x_i}=0$, Equation \ref{system2.1} becomes:
\begin{equation}
\begin{aligned}
0 & =(1-x_i^0)x_i^0\sum\limits_{j=1}^M \phi_{ij}(-1+\omega I_j)\sum\limits_{k=1}^M\phi_{kj} \\
\end{aligned}
\label{prep1}
\end{equation}
Equation (\ref{prep1}) leads to two solutions:  $x^0=0$ or $x^0= 1$.  \qed 

We can now obtain $R_0$ for the global dynamics in this model by using the Next Generation Approach, where $R_0$ is given by the most dominant eigenvalue (or `spatial radius' $\mathbf{\rho}$) of $\mathbf{F}\mathbf{V}^{-1}$, where $\mathbf{F}$ and $\mathbf{V}$ are $M \times M$ matrices, representing the `new infections' and `cases removed or transferred from the infected class', respectively in the disease free condition \cite{NGM1990,NGM2010}. As a result, $R_0$ is determined as follows (see Appendix \ref{app1} for detailed derivation):
\begin{equation}
R_0=\mathbf{\rho}(\mathbf{FV}^{-1}) 
\label{eq_SIR}
\end{equation} 
where 
\begin{equation}
\mathbf{F}=
\begin{bmatrix}\frac{\partial \mathcal{F}_1}{\partial I_1} & \frac{\partial \mathcal{F}_1}{\partial I_2} & \cdots & \frac{\partial \mathcal{F}_1}{\partial I_M} \\
\frac{\partial \mathcal{F}_2}{\partial I_1} & \frac{\partial \mathcal{F}_2}{\partial I_3} & \cdots & \frac{\partial \mathcal{F}_2}{\partial I_M} \\
\vdots                            & \vdots                            & \ddots & \vdots \\
\frac{\partial \mathcal{F}_M}{\partial I_1} & \frac{\partial \mathcal{F}_M}{\partial I_2} & \cdots & \frac{\partial \mathcal{F}_M}{\partial I_M} \\
\end{bmatrix}
\end{equation}	

\begin{equation}
\mathbf{V}=
\begin{bmatrix}\frac{\partial \mathcal{V}_1}{\partial I_1} & \frac{\partial \mathcal{V}_1}{\partial I_2} & \cdots & \frac{\partial \mathcal{V}_1}{\partial I_M} \\
\frac{\partial \mathcal{V}_2}{\partial I_1} & \frac{\partial \mathcal{V}_2}{\partial I_2} & \cdots & \frac{\partial \mathcal{V}_2}{\partial I_M} \\
\vdots                            & \vdots                            & \ddots & \vdots \\
\frac{\partial \mathcal{V}_M}{\partial I_1} & \frac{\partial \mathcal{V}_M}{\partial I_2} & \cdots & \frac{\partial \mathcal{V}_M}{\partial I_M} \\
\end{bmatrix}
\end{equation}
while
\begin{equation}
\begin{aligned}
\mathcal{F}_i &=S_i^0\sum_{M}^{j=1} \sum_{M}^{k=1}\beta_j\phi_{ij} \frac{\phi_{kj}I_k^0}{\epsilon^p_j} \\
\mathcal{V}_i &=\gamma I_i^0 +\mu I_i^0 \\
\end{aligned}
\end{equation}
Assuming $\beta_i=\beta$ for all nodes, we can derive $\mathbf{F}$ and $\mathbf{V}$ as follows:
\begin{equation}
\begin{aligned}
\mathbf{F} &= \beta 
\begin{bsmallmatrix}
\sum \limits_{j=1}^M S_1^0\frac{\phi_{1j} \phi_{1j}}{\epsilon^p_j} & \sum \limits_{j=1}^M S_1^0\frac{\phi_{1j} \phi_{2j}}{\epsilon^p_j} & \cdots &\sum \limits_{j=1}^M S_1^0\frac{\phi_{1j} \phi_{Mj}}{\epsilon^p_j}\\
\sum \limits_{j=1}^M S_2^0\frac{\phi_{2j} \phi_{1j}}{\epsilon^p_j} & \sum \limits_{j=1}^M S_2^0\frac{\phi_{2j} \phi_{2j}}{\epsilon^p_j} & \cdots &\sum \limits_{j=1}^M S_2^0\frac{\phi_{2j} \phi_{Mj}}{\epsilon^p_j} \\
\vdots & \vdots & \ddots & \vdots \\
\sum \limits_{j=1}^M S_M^0\frac{\phi_{Mj} \phi_{1j} }{\epsilon^p_j} & \sum \limits_{j=1}^M S_M^0\frac{\phi_{Mj} \phi_{2j}}{\epsilon^p_j} & \cdots &\sum \limits_{j=1}^M S_M^0\frac{\phi_{Mj} \phi_{Mj}}{\epsilon^p_j} \\
\end{bsmallmatrix} \\
\\
\mathbf{V} &=(\gamma+\mu) \mathbb{I} \\
\end{aligned}
\end{equation}
where $\mathbb{I}$ is a $M \times M$ identity matrix.\\

Using \textbf{Proposition 1}, $\mathbf{F}$ can be simplified as:
\begin{equation}
\mathbf{F}= \beta S^0
\begin{bsmallmatrix}
\sum \limits_{j=1}^M \frac{\phi_{1j} \phi_{1j}}{\epsilon^{p}_j} & \sum \limits_{j=1}^M \frac{\phi_{1j} \phi_{2j}}{\epsilon^{p}_j} & \cdots &\sum \limits_{j=1}^M \frac{\phi_{1j} \phi_{Mj}}{\epsilon^{p}_j}\\
\sum \limits_{j=1}^M \frac{\phi_{2j} \phi_{1j}}{\epsilon^{p}_j} & \sum \limits_{j=1}^M \frac{\phi_{2j} \phi_{2j}}{\epsilon^{p}_j} & \cdots &\sum \limits_{j=1}^M \frac{\phi_{2j} \phi_{Mj}}{\epsilon^{p}_j} \\
\vdots & \vdots & \ddots & \vdots \\
\sum \limits_{j=1}^M \frac{\phi_{Mj} \phi_{1j} }{\epsilon^{p}_j} & \sum \limits_{j=1}^M \frac{\phi_{Mj} \phi_{2j}}{\epsilon^{p}_j} & \cdots &\sum \limits_{j=1}^M \frac{\phi_{Mj} \phi_{Mj}}{\epsilon^{p}_j} \\
\end{bsmallmatrix}
\end{equation}

We can now prove the following simple but useful proposition.

\textbf{Proposition 2.} Matrix $\mathbf{G}$, defined as follows:
\begin{equation}
\mathbf{G}=  \begin{bsmallmatrix}
\sum \limits_{j=1}^M \frac{\phi_{1j} \phi_{1j}}{\epsilon^{p}_j} & \sum \limits_{j=1}^M \frac{\phi_{1j} \phi_{2j}}{\epsilon^{p}_j} & \cdots &\sum \limits_{j=1}^M \frac{\phi_{1j} \phi_{Mj}}{\epsilon^{p}_j}\\
\sum \limits_{j=1}^M \frac{\phi_{2j} \phi_{1j}}{\epsilon^{p}_j} & \sum \limits_{j=1}^M \frac{\phi_{2j} \phi_{2j}}{\epsilon^{p}_j} & \cdots &\sum \limits_{j=1}^M \frac{\phi_{2j} \phi_{Mj}}{\epsilon^{p}_j} \\
\vdots & \vdots & \ddots & \vdots \\
\sum \limits_{j=1}^M \frac{\phi_{Mj} \phi_{1j} }{N^{p^*}_j} & \sum \limits_{j=1}^M \frac{\phi_{Mj} \phi_{2j}}{N^{p^*}_j} & \cdots &\sum \limits_{j=1}^M \frac{\phi_{Mj} \phi_{Mj}}{N^{p^*}_j} \\
\end{bsmallmatrix}
\label{gmarkov}
\end{equation} 
is a Markov matrix.\\

\textit{Proof:} 
We assume all nodes have the same population $N$. Then  Equation (\ref{normalisation})
can be reduced to:
\begin{equation}
\epsilon ^{p^*}_i=\sum^{M}_{k=1}\phi_{ki}
\label{samepop}
\end{equation}
Without loss of generality, substituting Equation (\ref{samepop}) into Equation (\ref{gmarkov}) and expanding entries in the first column yields the first column $\mathbf{g_1}$ of matrix $\mathbf{G}$:

\begin{equation}
\mathbf{g_1}=\begin{bmatrix}
\frac{\phi_{11}\phi_{11}}{\sum \limits_{k=1}^M \phi_{k1}}+\frac{\phi_{12}\phi_{12}}{\sum \limits_{k=1}^M \phi_{k2}}+\cdots+\frac{\phi_{1M}\phi_{1M}}{\sum \limits_{k=1}^M \phi_{kM}} & \cdots \\

\frac{\phi_{21}\phi_{11}}{\sum \limits_{k=1}^M \phi_{k1}}+\frac{\phi_{22}\phi_{12}}{\sum \limits_{k=1}^M \phi_{k2}}+\cdots+\frac{\phi_{2M}\phi_{1M}}{\sum \limits_{k=1}^M \phi_{kM}} & \cdots  \\

\vdots & \vdots   \\

\frac{\phi_{M1}\phi_{11}}{\sum \limits_{k=1}^M \phi_{k1}}+\frac{\phi_{M2}\phi_{12}}{\sum \limits_{k=1}^M \phi_{k2}}+\cdots+\frac{\phi_{MM}\phi_{1M}}{\sum \limits_{k=1}^M \phi_{kM}} & \cdots \\
\end{bmatrix}
\end{equation}
Continuing with column 1, the column sum is: 
\begin{equation}
\begin{aligned}
& =\phi_{11}\frac{\sum \limits_{k=1}^M\phi_{k1}}{\sum \limits_{k=1}^M \phi_{k1}}+\phi_{12}\frac{\sum \limits_{k=1}^M\phi_{k2}}{\sum \limits_{k=1}^M \phi_{k2}}+\cdots+\phi_{1M}\frac{\sum \limits_{k=1}^M\phi_{kM}}{\sum \limits_{k=1}^M \phi_{kM}} \\
& =\sum \limits_{k=1}^M\phi_{1k}  =1 \\
\end{aligned}
\end{equation}
Similarly, the sums of each column are equal to 1, with all entries being non-negative population fractions. Hence, $\mathbf{G}$  is a Markov matrix. \qed 

As a Markov matrix, $\mathbf{G}$ always has the most dominant eigenvalue of unity. All other eigenvalues are smaller than unity in absolute value \cite{markov_matrix}.

Now the next generation matrix $\mathbf{K}$ can be obtained as follows:
\begin{equation}
\begin{aligned}
\mathbf{K} & = \frac{\beta S^0}{\gamma+\mu} \mathbf{G} \\
& = \frac{\beta[(1-x^0)+\zeta x^0]}{\gamma+\mu} \mathbf{G} \\
\label{K_SIR_newborn}
\end{aligned}
\end{equation}

From \textbf{Proposition 2} and Equation (\ref{eq_SIR}), $R_0$ can be obtained as: 
\begin{equation} \label{r0_exp}
\begin{aligned}
R_0 &=\rho(\mathbf{K}) \\
&=\frac{\beta[(1-x^0)+\zeta x^0]}{\gamma+\mu} \rho(\mathbf{G}) \\
&=\frac{\beta[(1-x^0)+\zeta x^0]}{\gamma+\mu} \\
\end{aligned}
\end{equation}

Noting Proposition 1, there are two cases: $x=0$ and $x=1$.  When nobody vaccinates ($x^0=0, S^0=1$), the entire population remains susceptible, which reduces model (\ref{system2.1}) to a canonical SIR model without vaccination intervention and $R_0$ returns to $\frac{\beta}{\gamma+\mu}$. On the other hand, if the whole population is vaccinated ($x^0=1$), the fraction of susceptible population only depends on the vaccine failure rate $\zeta$, resulting in:
\begin{equation}
R_0=\frac{\beta \zeta}{\gamma+\mu}
\label{R0_2.1}
\end{equation}
Clearly, fully effective vaccination ($\zeta=0$) would prohibit disease spread ($R_0=0$). Partially effective vaccination could potentially suppress disease transmission, or even eradicate disease spread if $R_0<1$. If all vaccinations fail ($\zeta=1$), model \ref{system2.1} concurs with a canonical SIR model without vaccination, which also corresponds to a special case in Equation \ref{Adino_R0} where $\epsilon \to \infty$.

\subsection{Vaccination available to the entire susceptible class}

If the vaccination opportunity is expanded to the entire susceptible class, including newborns and adults, the following model is proposed based on model (\ref{system2.1}): 
\footnotesize
\begin{equation}
\begin{aligned}
\dot{S_i} &=\mu-\sum\limits_{j=1}^M \sum\limits_{k=1}^{M} \beta_j \phi_{ij} \frac{\phi_{kj}I_k}{\epsilon^p_j} S_i-\mu S_i- x_i S_i+x_i \zeta S_i  \\
\dot{I_i}  &=\sum\limits_{j=1}^M \sum\limits_{k=1}^{M} \beta_j \phi_{ij} \frac{\phi_{kj}I_k}{\epsilon^p_j}S_i -\gamma I_i -\mu I_i \\
\dot{R_i} &=  S_i x_i (1-\zeta)+\gamma I_i- \mu I_i \\
\dot{x_i} &=\kappa (1-x_i)\sum\limits_{j=1}^M \phi_{ij}(-1+\omega I_j)\sum\limits_{k=1}^M\phi_{kj}x_k \\
\end{aligned}
\label{system2.2}
\end{equation} 
\normalsize

Model (\ref{system2.2}) has a disease-free equilibrium $(S^0_i, I^0_i, x^0_i)$: 

\begin{equation}
\begin{aligned} 
S_i^0 & =\frac{\mu}{\mu+x_i^0(1-\zeta)}\\
I_i^0 & =0 \\ 
\end{aligned}
\label{eq_sus_equ}
\end{equation}
while
\begin{equation}
\begin{aligned}
\mathcal{F}_i &=S_i^0\sum_{j=1}^{M} \sum_{k=1}^{M}\beta_j\phi_{ij} \frac{\phi_{kj}I_k^0}{\epsilon^p_j} \\
\mathcal{V}_i &=\gamma I_i^0 +\mu I_i^0 \\
\end{aligned}
\end{equation}
Substituting $S^0_i$ using Equation (\ref{eq_SIR2.2}) yields
\begin{equation}
\begin{aligned}
\mathbf{F} &= \beta \frac{\mu}{\mu+x^0(1-\zeta)} \mathbf{G} \\
\mathbf{V} &=(\gamma+\mu) \mathbb{I} \\
\end{aligned}
\end{equation}
where $\mathbb{I}$ is a $M \times M$ identity matrix.

The next generation matrix $\mathbf{K}$ can then be obtained as:
\begin{equation}
\mathbf{K} = \frac{\beta \mu}{(\gamma+\mu)[\mu+x^0(1-\zeta)]} \mathbf{G}  \\
\label{K_SIR_s}
\end{equation}
Using  \textbf{Proposition 2}, $R_0$ can be obtained as:
\begin{equation}
R_0= \frac{\beta \mu}{(\gamma+\mu)[\mu+x^0(1-\zeta)]}
\label{R0_2.2}
\end{equation}

When nobody vaccinates ($x^0=0$), model (\ref{system2.2}) concurs with a canonical SIR model with $R_0$ reducing to $\frac{\beta}{\gamma+\mu}$. When the vaccination attains the full coverage in the population ($x^0=1$), the magnitude of $R_0$ depends on the vaccine failure rate $\zeta$:
\begin{equation}
R_0=\frac{\beta \mu}{(\gamma+\mu)(\mu+1-\zeta)}
\label{R0_2.2_x1}
\end{equation} 

Equation \ref{R0_2.2_x1} reduces to $\frac{\beta}{\gamma+\mu}$ if all vaccines fail ($\zeta=1$). Conversely, if all vaccines are effective $\zeta=0$, Equation \ref{R0_2.2_x1} yields
\begin{equation}
R_0 =\frac{\beta \mu}{(\gamma+\mu)(\mu+1)}
\end{equation}
Given that $\mu \ll 1$, $R_0$ is well below the critical threshold:
\begin{equation}
R_0=\frac{\beta \mu}{\gamma} \ll 1
\label{R0_2.2_x1z0}
\end{equation}

If $0<\zeta<1$, by comparing Equations (\ref{R0_2.1}) and (\ref{R0_2.2_x1}), we note that $R_0$ becomes smaller, provided $ \zeta > \mu$, that is: 
\begin{equation}
\begin{aligned}
R_{0}^{newborn}-R_0^{susceptible} &=\frac{\beta \zeta}{\gamma+\mu}-\frac{\beta \mu}{(\gamma+\mu)(\mu+1-\zeta)} \\
&=\frac{\beta (1-\zeta)(\zeta-\mu)}{(\gamma+\mu)(\mu+1-\zeta)}  >0 \\                                                                	
\end{aligned}
\end{equation}

\subsection{Vaccination available to the entire susceptible class with committed vaccine recipients}

We now consider the existence of committed vaccine recipients, $x^c$, as a fraction of individuals who would choose to vaccinate regardless of payoff assessment \cite{Bai2015,SEIR2013} ($0 < x^c \ll 1$). We assume that committed vaccine recipients are also exposed to vaccination failure rate $\zeta$ and are distributed uniformly across all nodes. It is also important to point out that the fraction of committed vaccine recipients is  constant over time. However, they can still affect vaccination decision for those who are not vaccinated, and consequently, contribute to the rate of change of the vaccinated fraction $x$. Model (\ref{system2.2}) can be further extended to reflect these considerations, as follows:
\footnotesize
\begin{equation}
\begin{aligned}
\dot{S_i} &=\mu-\sum\limits_{j=1}^M \sum\limits_{k=1}^{M} \beta_j \phi_{ij} \frac{\phi_{kj}I_k}{\epsilon ^p_j} S_i-\mu S_i+(\zeta-1) [(x_i+x^c) S_i]  \\
\dot{I_i} &=\sum\limits_{j=1}^M \sum\limits_{k=1}^{M} \beta_j \phi_{ij} \frac{\phi_{kj}I_k}{\epsilon ^p_j}S_i -\gamma I_i -\mu I_i \\
\dot{x_i} &=\kappa (1-x_i-x^c)\sum\limits_{j=1}^M \phi_{ij}(-1+\omega I_j)\sum\limits_{k=1}^M\phi_{kj}(x_k+x^c) \\
\end{aligned}
\label{system_reform}
\end{equation} 
\normalsize

Model (\ref{system_reform}) has a disease-free equilibrium:
\begin{equation}
\begin{aligned} 
S_i^0 & =\frac{\mu}{\mu+(x_i^0+x^c)(1-\zeta)}\\
I_i^0 & =0 \\ 
\end{aligned}
\label{eq_SIR3}
\end{equation}
while
\begin{equation}
\begin{aligned}
\mathcal{F}_i &=S_i^0 \sum_{M}^{j=1} \sum_{M}^{k=1}\beta_j\phi_{ij} \frac{\phi_{kj}I_k^0}{\epsilon ^p_j} \\
\mathcal{V}_i &=\gamma I_i^0 +\mu I_i^0 \\
\end{aligned}
\end{equation}
Substituting $S^0_i$, using Equation \ref{eq_SIR3}, yields:
\begin{equation}
\begin{aligned}
\mathbf{F} &= \frac{\beta \mu}{\mu+(x^0+x^c)(1-\zeta)}
 \begin{bsmallmatrix} 
\sum \limits_{j=1}^M \frac{\phi_{1j} \phi_{1j}}{\epsilon ^p_j} & \sum \limits_{j=1}^M \frac{\phi_{1j} \phi_{2j}}{\epsilon ^p_j} & \cdots &\sum \limits_{j=1}^M \frac{\phi_{1j} \phi_{Mj}}{\epsilon ^p_j}\\
\sum \limits_{j=1}^M \frac{\phi_{2j} \phi_{1j}}{\epsilon ^p_j} & \sum \limits_{j=1}^M \frac{\phi_{2j} \phi_{2j}}{\epsilon ^p_j} & \cdots &\sum \limits_{j=1}^M \frac{\phi_{2j} \phi_{Mj}}{\epsilon ^p_j} \\
\vdots & \vdots & \ddots & \vdots \\
\sum \limits_{j=1}^M \frac{\phi_{Mj} \phi_{1j} }{\epsilon ^p_j} & \sum \limits_{j=1}^M \frac{\phi_{Mj} \phi_{2j}}{\epsilon ^p_j} & \cdots &\sum \limits_{j=1}^M \frac{\phi_{Mj} \phi_{Mj}}{\epsilon ^p_j} \\
\end{bsmallmatrix} \\
\mathbf{V} &=(\gamma+\mu) \mathbb{I} \\
\end{aligned}
\end{equation}
where $\mathbb{I}$ is a $M \times M$ identity matrix.\\

The next generation matrix $\mathbf{K}$ can now be obtained as:
\begin{equation}
\mathbf{K} = \frac{\beta \mu}{[\mu+(x^0+x^c)(1-\zeta)](\gamma+\mu)}  \mathbf{G} \\
\label{K_SIR_com}
\end{equation}

\textbf{Proposition 3}  At the disease free equilibrium, $x^0$ has two solutions $x^0=x^c=0$ or $x^0+x^c=1$ if $x^0$, $x^c$ and $S^0$ are uniform across all nodes.

\textit{Proof:}
In analogy with \textbf{Proposition 1}, with committed vaccine recipients, at the disease free equilibrium, $S_i^0$ and $x_i^0$ are uniform across all nodes, and therefore:
\begin{equation}
S^0_i =  S^0, \ x_i^0 = x^0+x^c \quad \forall i
\end{equation}
Hence, under disease-free condition where $I^0=0$ and $\dot{x_i}=0$, Equation \ref{system_reform} becomes:
\begin{equation}
\begin{aligned}
0 & =(1-x_i^0-x^c)(x_i^0+x^c)\sum\limits_{j=1}^M \phi_{ij}(-1+\omega I_j)\sum\limits_{k=1}^M\phi_{kj} \\
\end{aligned}
\label{prop3}
\end{equation}
Equation \ref{prop3} leads to two solutions:  $x^0+x^c=0$ or $x^0+x^c=1$.  \qed

$R_0$ can be obtained for this case as:
\begin{equation}
R_0=\frac{\beta \mu}{[\mu+(x^0+x^c)(1-\zeta)](\gamma+\mu)} 
\label{R0_3.2}
\end{equation}
 If nobody chooses to vaccinate  ($x^0=x^c=0$), $R_0$ can be reduced to $\frac{\beta}{\gamma+\mu}$.
Conversely, if the entire population is vaccinated $(x^0+x_c=1)$, Equation (\ref{R0_3.2}) reduces to equation (\ref{R0_2.2_x1}), and $R_0$ is purely dependent on the vaccine failure rate $\zeta$.

\subsection{Model parametrization}
The proposed models aim to simulate a scenario of a generic childhood disease (e.g., measles) where life-long full immunity is acquired after effective vaccination. The vaccination failure rate, $\zeta$, is set as the probability of ineffective vaccination, to showcase the influence of unsuccessful vaccination on the global epidemic dynamics. In reality, the vaccination for Measles-Mumps-Rubella  (MMR) is highly effective: for example, in Australia, an estimated 96\% is  successful in conferring immunity \cite{vacc_failure}. 

The population flux matrix $\phi$ is derived from the network topology, in which each entry represents the connectivity between two nodes: if two nodes are not connected, $\phi_{ij}=0$; if two nodes are connected, $\phi_{ij}$ is randomly assigned in the range of $(0,1]$. The population influx into a node $i$ within a day is represented by the column sum $\sum_{j=1}^M \phi_{ji}$ in flux matrix $\phi$. 

The parameters used for all simulations are summarized in Table \ref{parameter}. Same initial conditions are applied to all nodes. Parameters $\omega$ and $\kappa$ are calibrated based on values used in \cite{bauch_2005_imi}. \\

\begin{table*}[tbh!]
	\renewcommand{\arraystretch}{1.2}
	\begin{tabular}{llll}
		\hline \noalign{\smallskip}
		parameter  & Interpretation                                  & Baseline value & References \\
		\noalign{\smallskip}\hline\noalign{\smallskip}
		$1/\gamma$ & Average length of recovery period (days)        & 10             & \cite{measle_R0}    \\
		$R_0$      & Basic reproduction number                       & 15             & \cite{measle_R0}\\ 
		$\mu$      & Mean birth and death rate ($days^{-1}$)         & 0.000055       & \cite{bauch_2005_imi}      \\
		$\zeta$    & Vaccination failure rate                        & [0,1]          & Assumed            \\
		$\kappa$   & Imitation rate                                  & 0.001          & \cite{bauch_2005_imi}      \\
		$\omega$   & Responsiveness to changes in disease prevalence & [1000,3500]    & \cite{bauch_2005_imi}     \\
		$\phi_{ij}$& Fraction of residents from node $i$ travelling to $j$ & [0,1]    & Network connectivity \\
		$I$      & Initial condition                               & 0.001          & \cite{bauch_2005_imi}  \\
		$S$      & Initial condition                               & 0.05           & \cite{bauch_2005_imi}  \\
		$x$      & Initial condition                               & 0.95           & \cite{bauch_2005_imi}  \\
		\noalign{\smallskip}\hline  
	\end{tabular}
	\caption{Epidemiological and behavioural parameters}
	\label{parameter}
\end{table*}

Our simulations were carried out on two networks: 
\begin{itemize}
	\item a pilot case of a network with 3 nodes (suburbs), and 
	\item an Erd\"os-R\'enyi random network \cite{random_network} with 3000 nodes (suburbs). 
\end{itemize}
Each of these networks was used in conjunction with the following three  models of vaccination behaviours:
\begin{itemize}
	\item vaccinating newborns only: using model (\ref{system2.1}) 
	\item vaccinating the susceptible class: using model (\ref{system2.2}), 
	\item vaccinating the susceptible class with committed vaccine recipients: using model (\ref{system_reform}). 
\end{itemize}

\begin{figure*}[htb!]
	\captionsetup[subfigure]{justification=centering}
	\subfloat[No population mobility.\newline 
	$i=j=\{1,2,3\},\phi_{ij}=0$ where $i \neq j$. Otherwise $\phi_{ij}=1$.]{
		\includegraphics[width=0.5\textwidth,keepaspectratio]{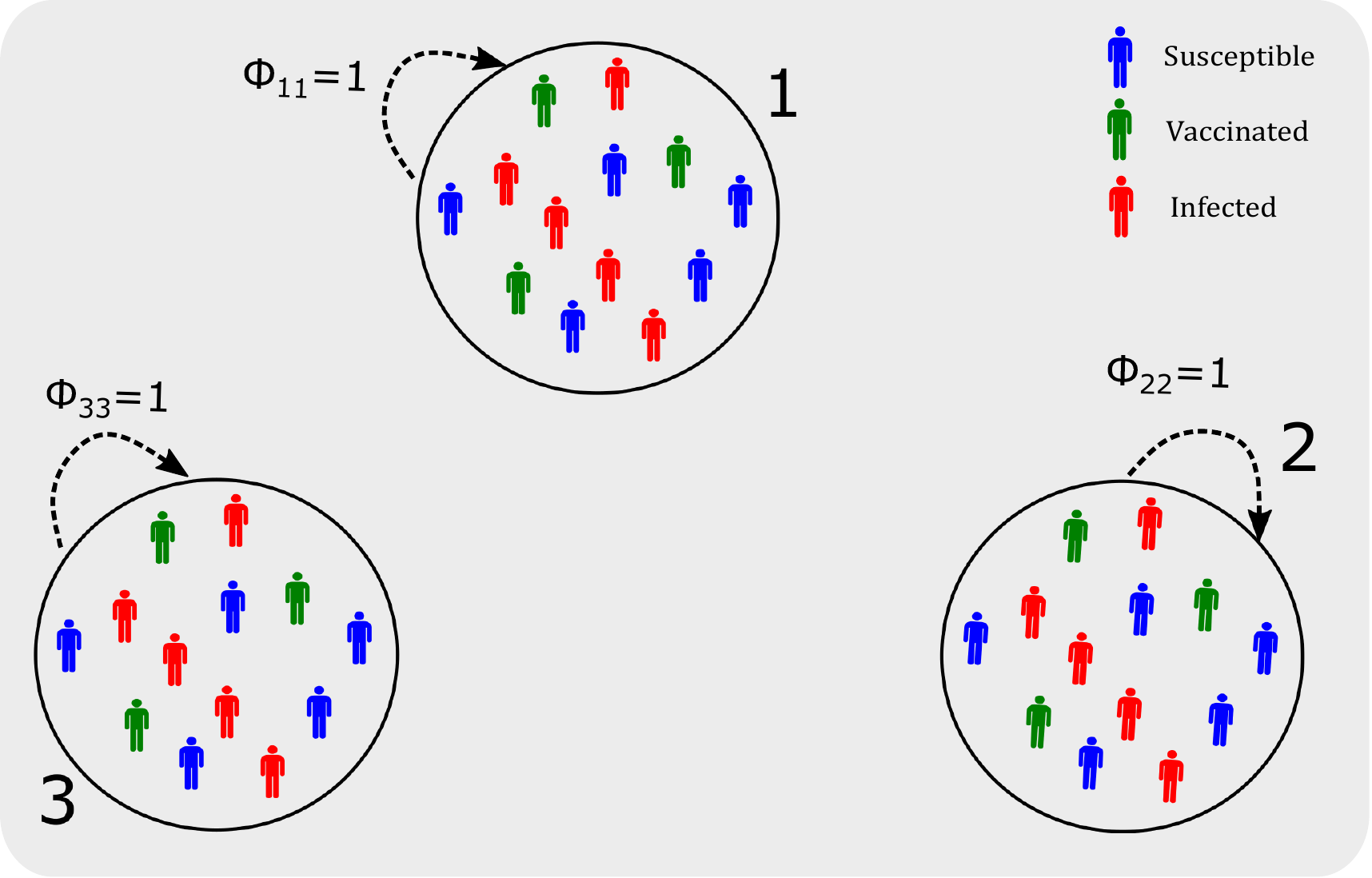}}
	\subfloat[Equal population mobility.\newline 
	$i=j=\{1,2,3\},\phi_{ij}=\frac{1}{3}$]{
		\includegraphics[width=0.5\textwidth,keepaspectratio]{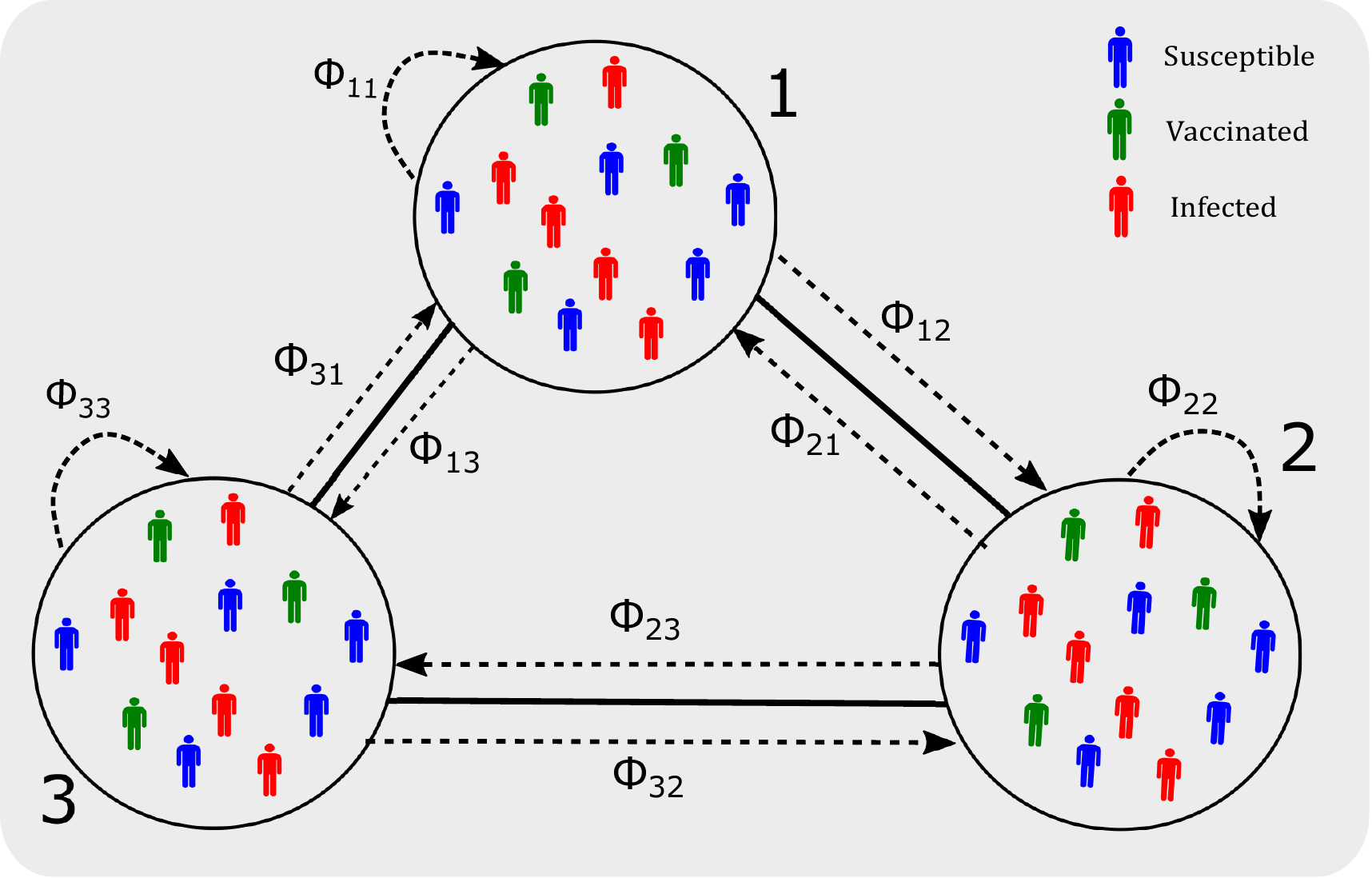}}
	\caption{Schematic of the 3-node case: population mobility across nodes }
	\label{3-nodeSche}	
\end{figure*}

In the pilot case of 3-node network, two network topologies are studied: an isolated 3-node network where all residents remain in their residential nodes without travelling to other nodes (equivalent to models proposed in \cite{bauch_2005_imi}), and a fully connected network where the residents at each node commute to the other two nodes,  and the population fractions commuting are symmetric and uniformly distributed (Figure \ref{3-nodeSche}). 

We then consider a suburb-network modelled as an  Erd\"os-R\'enyi random graph  (with number of nodes $M=3000$) with average degree $\langle k \rangle =4$, in order to study how an expansive travelling pattern affects the global epidemic dynamics. Other topologies, such as lattice, scale-free and small-world networks, can easily be substituted here, though in this study our focus is on an Erd\"os-R\'enyi   random graphs. 

\section{Results}
\label{results}
\subsection{3-node network}

\subsubsection{Vaccinating newborns only}

\begin{figure*}[htb!]
	\centering
	\graphicspath{{./fig/}}
	\includegraphics[width=\textwidth]{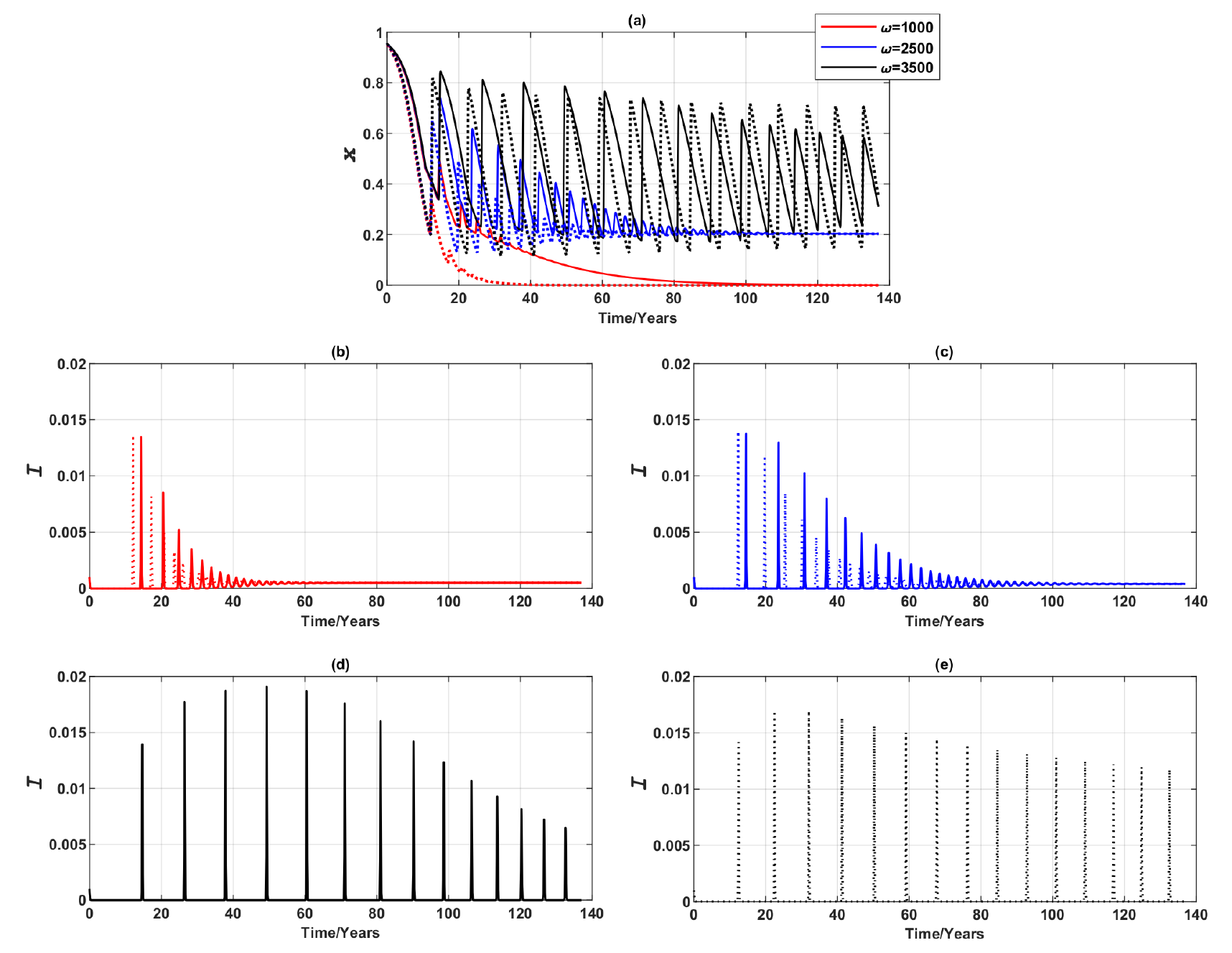}
	\caption{Epidemic dynamics of a 3-node case for three values of $\omega$ when vaccinating newborns. Time series of (a)  proportion of vaccinated individuals, \textit{x}, and (b)-(e) Infection prevalence, \textit{I}.  Solid line: Symmetric uniform population mobility. Dotted line: No population mobility. Commuting suppresses prevalence peaks over time at high $\omega$, but may produce higher prevalence peaks over time at mid and low $\omega$.}
	\label{3-node}	
\end{figure*}

When there is no population mobility, we observe three distinctive equilibria (Figure \ref{3-node}; dotted lines): a pure non-vaccinating equilibrium (where $x^{f}=0$, representing the final condition, and $\omega=1000$), a mixed equilibrium (where $x^{f} \neq 0$, $\omega=2500$), and stable limit cycles (where $x^{f} \neq 0$,  $\omega=3500$). This observation is in qualitative agreement with the vaccinating dynamics reported by \cite{bauch_2005_imi}.  

When commuting is allowed, individuals commute to different nodes, and their decision will no longer rely on the single source of information (i.e., the disease prevalence in their residential node) but will also depend on the disease prevalence at their destination. As a consequence, the three distinctive equilibria are affected in different ways (Figure \ref{3-node}; solid lines). The amplitude of the stable limit cycles at   $\omega =  3500$ is reduced as a result of the reduced disease prevalence. It takes comparatively longer (compare the dotted lines and solid lines in Figure \ref{3-node}) to converge to the pure non-vaccinating equilibrium at $\omega=1000$, and to the endemic, mixed equilibrium at $\omega=2500$ with high amplitude of oscillation at the start of the epidemic spread. As $\omega$ can be interpreted as the responsiveness of vaccinating behaviour to the disease prevalence \cite{bauch_2005_imi}, if individuals are sufficiently responsive (i.e., $\omega$ is high), the overall epidemic is suppressed more due to population mobility (and the ensuing imitation), as evidenced by smaller prevalence peaks. Conversely, if individuals are insensitive to the prevalence change (i.e., $\omega$ is low), epidemic dynamics with equal population mobility may appear to be more volatile at the start, but the converged levels of both prevalence peak and vaccine uptake remain unchanged in comparison to the case where there is no population mobility.

\begin{figure*}[h!]
	\centering
	\includegraphics[width=\textwidth]{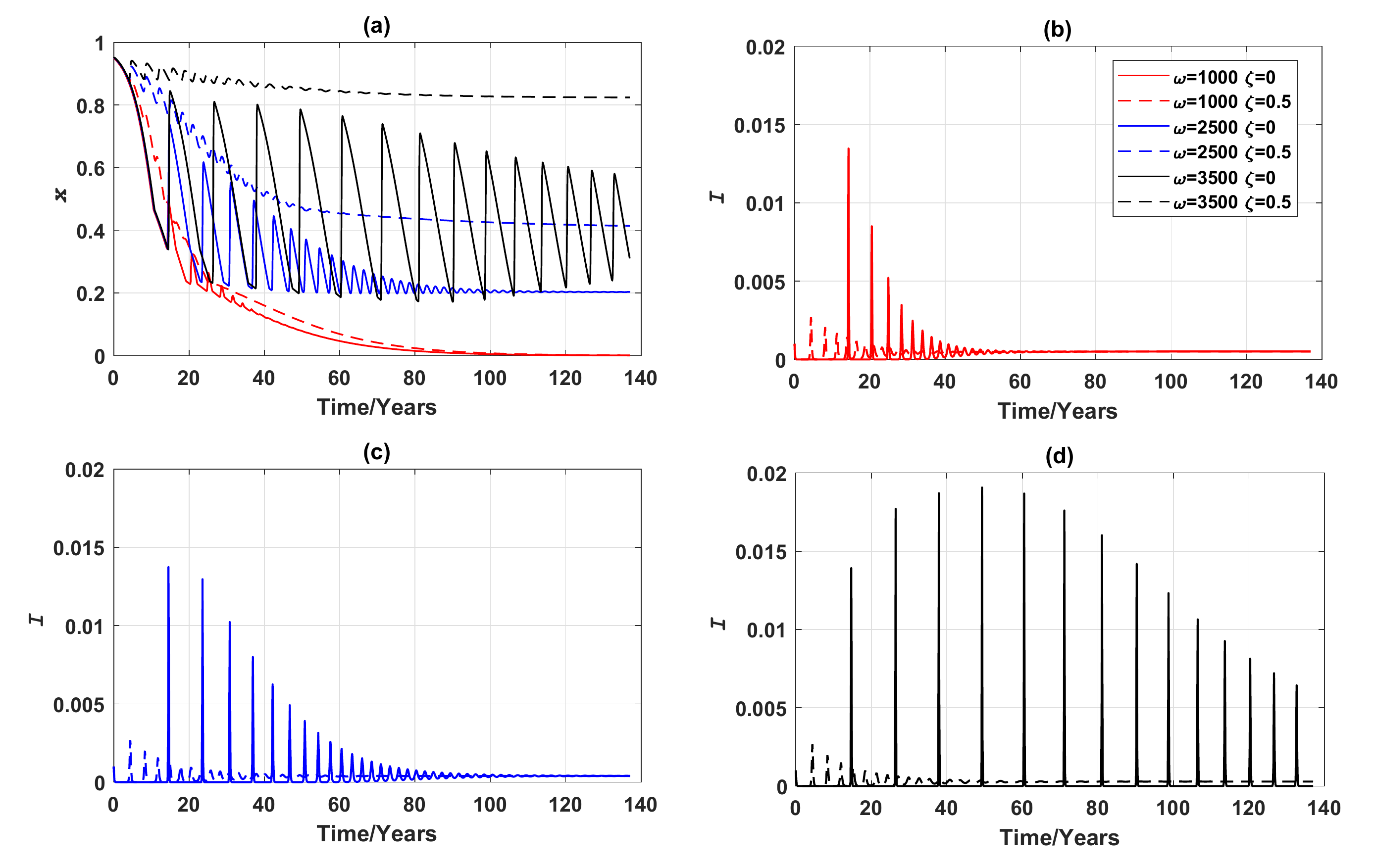}
	\caption{Comparison of vaccination failure rates: epidemic dynamics of a 3-node case for three values of $\omega$ (which measures the responsiveness of individuals to prevalence) when vaccinating newborns. (a)  Proportion of vaccinated individuals, \textit{x}, and (b)-(d) Proportion of infected individuals, \textit{I}. Solid line: $\zeta=0$. Dashed line: $\zeta=0.5$.}
	\label{3node_zeta}
\end{figure*}

We further investigated how the vaccination failure rate $\zeta$ affects the global epidemic dynamics. If, for example, only a half of the vaccine administered is effective ($\zeta=0.5$), as shown in Figure \ref{3node_zeta} (b-d),  the infection peaks arrive sooner, for all values of $\omega$. As a result of having the earlier infection peaks, the individuals respond to the breakout and choose to vaccinate sooner, causing the vaccine uptake to rise. This seemingly counter-intuitive behaviour has also been reported in \cite{Alessio2}. When $\zeta=0$, the prevalence peaks take longer to develop and the extended period gives individuals an illusion that there may not be an epidemic breakout, and consequently encourages `free-riding' behaviour. In the case where half of the vaccinations fail, epidemic breaks out significantly earlier at lower peaks, an observation that is beneficial to encourage responsive individuals to choose to vaccinate. Although some (in this case half) of the vaccines fail, a sufficiently high vaccine uptake still curbs prevalence peaks and shortens the length of breakout period. In terms of the final vaccination uptake, the vaccine failure rate predominantly affects behaviours of responsive individuals (when  $\omega$ is high, such as $3500$). In this case, instead of the stable limit cycles observed when $\zeta=0$, an endemic, mixed equilibrium is reached when the vaccination failure rate is significant. Final vaccination uptake is impacted little by vaccination failure rate when individuals are  insensitive to changes in disease prevalence (when $\omega$ is lower in value, such as $2500$ or $1000$). These observations are illustrated in Figure \ref{3node_zeta}.

\subsubsection{Vaccinating the entire susceptible class}
\begin{figure*}[htb!]
	\centering
	\graphicspath{{./fig/}}
	\includegraphics[width=\textwidth]{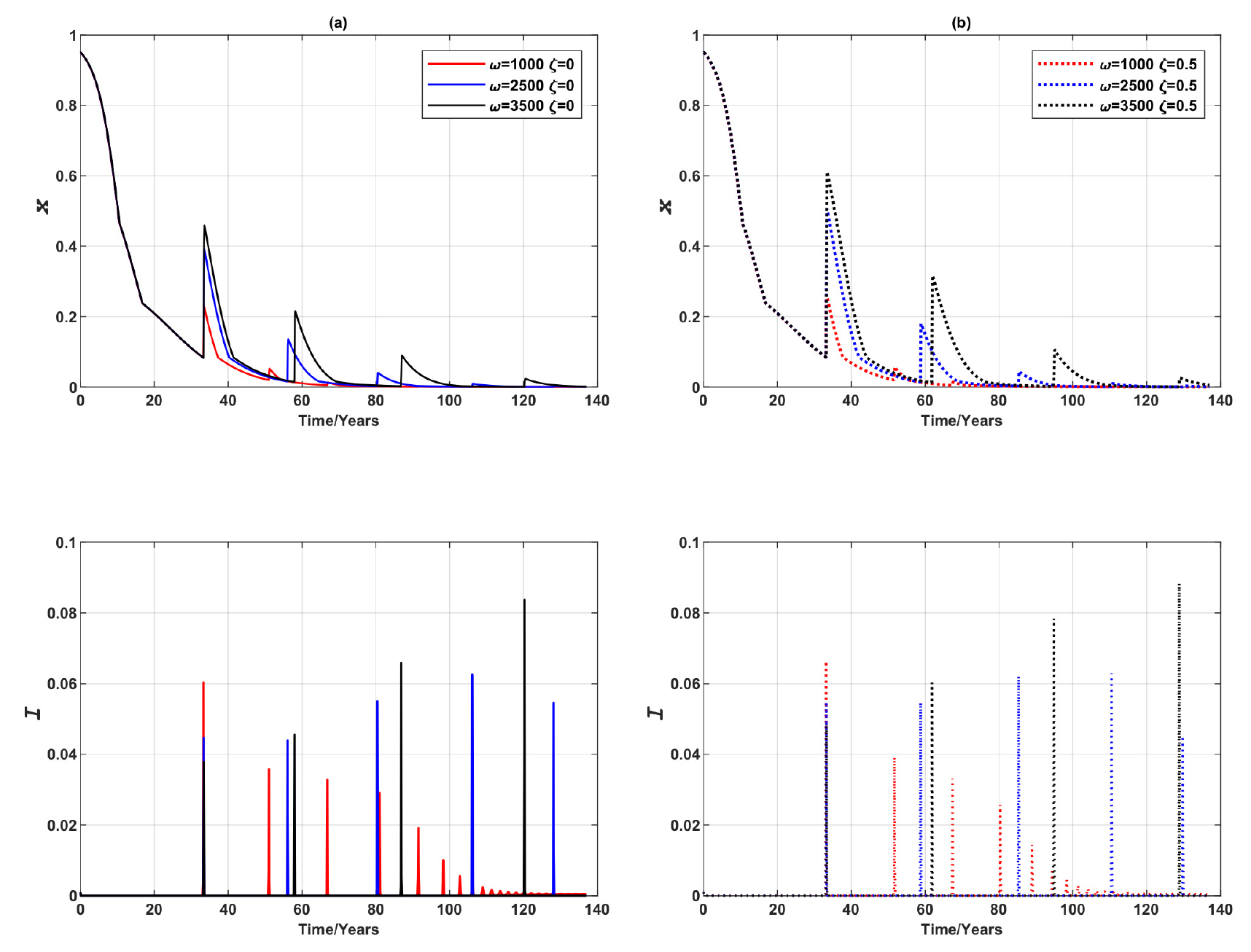}
	\caption{Comparison of vaccination failure rates: epidemic dynamics of a 3-node case for three values of $\omega$ (which measures the responsiveness of individuals to prevalence) when vaccinating newborns and adults. (a)-(b)  Proportion of vaccinated individuals, \textit{x}, and (c)-(d) Proportion of infected individuals, \textit{I}. Solid line: $\zeta=0$. Dashed line: $\zeta=0.5$.  }
	\label{3node_sus_zeta}	
\end{figure*}

If (voluntary) vaccination is offered to the susceptible class regardless of the age, the initial condition of $x=0.95$ represents the scenario that the vast majority of population are immunized to begin with, and the epidemic would not breakout until the false sense of security provided by the temporary `herd immunity' settles in. This feature is observed in Figure \ref{3node_sus_zeta} (a)-(b): the vaccination coverage continues to drop at the start of the epidemic breakout, indicating that individuals, regardless of their responsiveness towards the prevalence change, exploit the temporary herd immunity until an infection peak emerges. Since vaccination is available to the entire susceptible class, a small increase in $\textit{x}$ could help suppresses disease prevalence. However, when vaccination is partially effective (i.e., $\zeta=0.5$), the peaks in vaccine uptake, $\textit{x}$, are no longer an accurate reflection of the actual vaccination coverage, and such peaks therefore may not be sufficient to adequately suppress infection peaks. It can  also be observed that vaccinating the susceptible class encourages non-vaccinating behaviours due to the perception of herd immunity,  and therefore pushes the epidemic towards an endemic equilibrium, particularly when sensitivity to prevalence is relatively high ( mid and high $\omega$), as the responsive individuals would react promptly to the level of disease prevalence, altering their vaccination decisions. However,  no substantial impact is observed on those individuals who are insensitive to changes in the disease prevalence ( low $\omega$).
\subsection{Erd\"os-R\'enyi random network of 3000 nodes}

We now present the simulation results on a much larger  Erd\"o-R\'enyi random network of $M=3000$ nodes, which more realistically reflects the size of a modern city and its commuting patterns. It was observed by previous studies \cite{immunization} that such a larger system requires a higher vaccination coverage to achieve herd immunity, and thus curbs `free-riding' behaviours more effectively. 

\subsubsection{Vaccinating newborns only}

\begin{figure*}[htb!]
	\centering
	\graphicspath{{./fig/}}
	\includegraphics[width=\columnwidth,trim={2cm 0.5cm 2cm 0.5cm},clip]{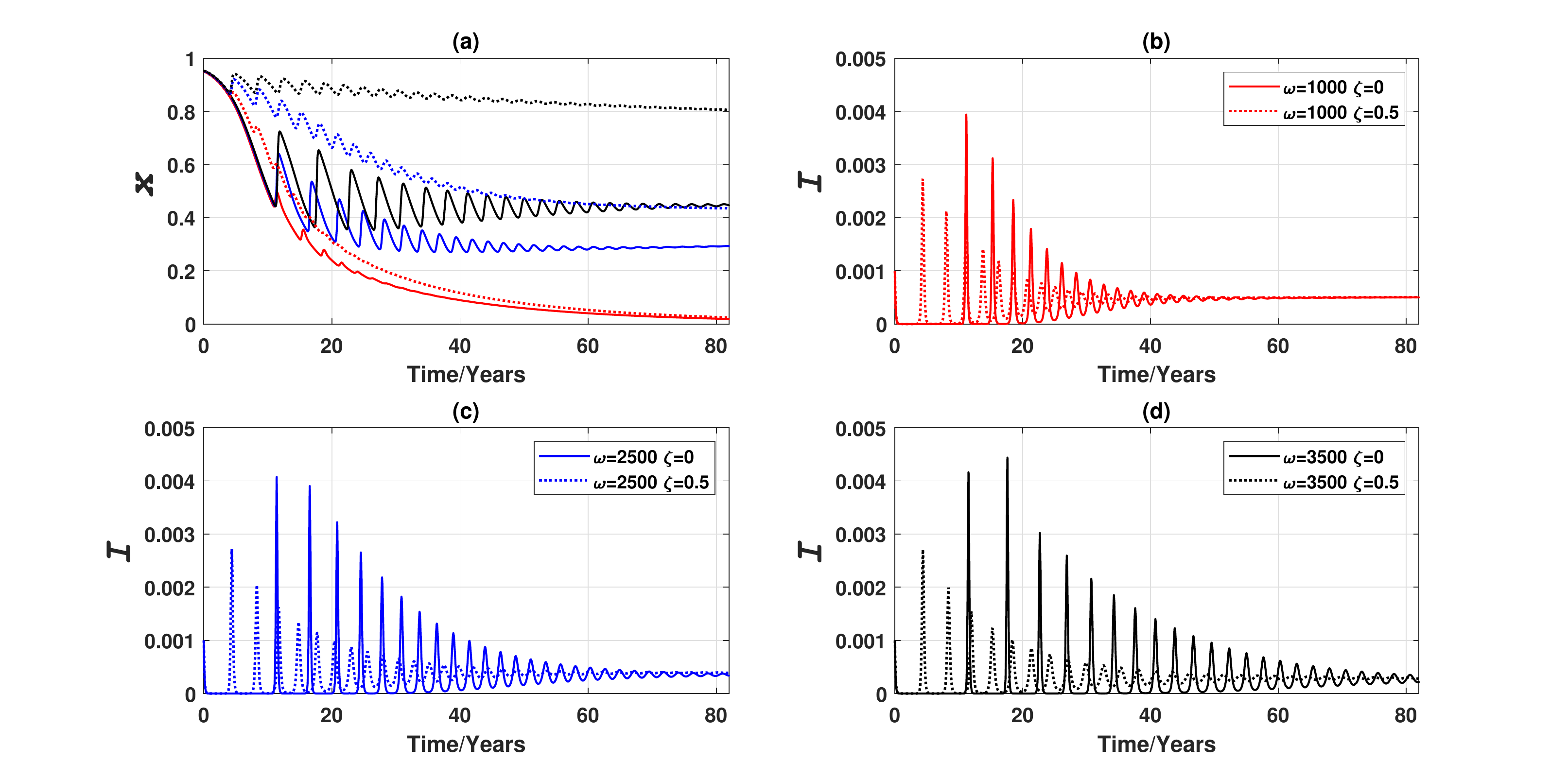}	
	\caption{Comparison of vaccination failure rates: epidemic dynamics of a Erd\"os-R\'enyi random network of 3000 nodes for three values of $\omega$ (which measures the responsiveness of individuals to prevalence) when vaccinating newborns. (a)  Proportion of vaccinated individuals, \textit{x}, and (b)-(d) Proportion of infected individuals, \textit{I}. Solid line: $\zeta=0$. Dashed line: $\zeta=0.5$.}	
	\label{newborn_zeta0}
\end{figure*}

In this case, three distinct equilibria are observed (as shown in Figure \ref{newborn_zeta0} (a)) for three values of $\omega$: a pure non-vaccinating equilibrium at $\omega=1000$ and two endemic mixed equilibria at $\omega=2500$ and $\omega=3500$ respectively, replacing the stable limit cycles at high $\omega$ previously observed in the 3-node case. As expected, when individuals are more responsive to disease prevalence (higher $\omega$), they are more likely to get vaccinated.  The expansive travelling pattern also somewhat elevates the global vaccination coverage level (compared to the 3-node case), particularly in the case that the individuals show a moderate level of responsiveness to prevalence ($\omega =  2500$), and shortens the convergence time to reach the equilibrium. However, for those who are insensitive to the changes in disease prevalence ($\omega =  1000$), the more expansive commuting presented by the larger network does not affect either the level of voluntary vaccination, or the convergence time in global dynamics -  these individuals remain unvaccinated as in the case of the smaller network. It is also found that the vaccination coverage is very sensitive to the disease prevalence change at the larger network since the disease prevalence peaks are notably lower than the 3-node case counterparts. If half of the vaccines administered are unsuccessful ($\zeta=0.5$), the impact of early peaks on global dynamics is magnified in a larger network (Figure \ref{newborn_zeta0} (b-d)), leading to a shorter convergence time, although the final equilibria are hardly affected compared to the 3-node case as shown in figure \ref{3node_zeta}. 

\begin{figure*}[htb!]
	\centering
	\graphicspath{{./fig/}}
	\includegraphics[width=\textwidth]{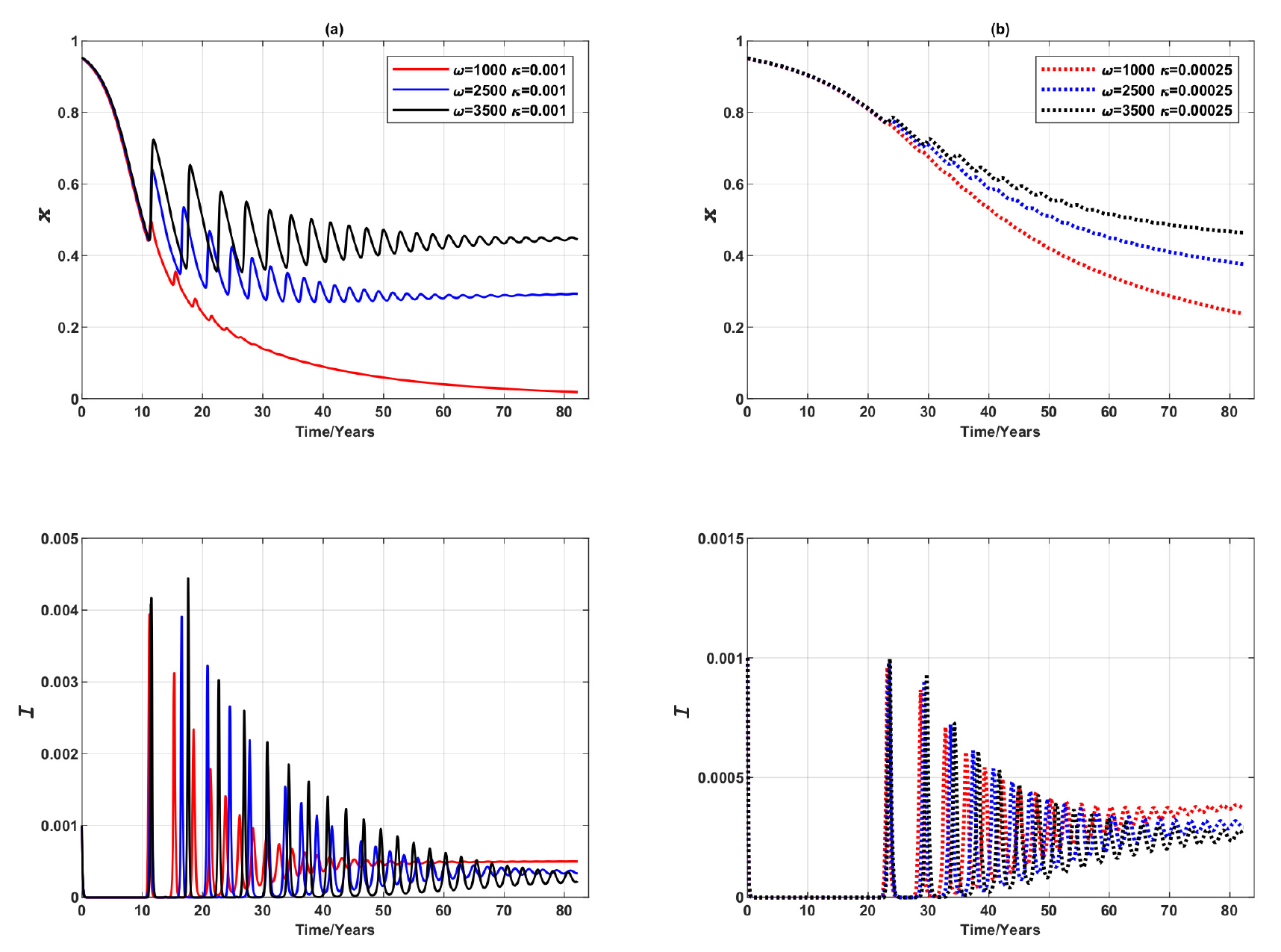}	
	\caption{Epidemic dynamics of a Erd\"os-R\'enyi random network of 3000 nodes,  varying the value of $\kappa$ for three values of $\omega$ (vaccinating newborn). Time series of $x$, frequency of vaccinated individuals, and $I$, Disease prevalence. (a) Solid line: $\kappa=0.001$. (b) Dotted line: $\kappa=0.00025$. }	
	\label{kappa_XI}
\end{figure*}

We also compared the epidemic dynamics in terms of the adjusted imitation rate, represented by the parameter $\kappa$, as shown in figure \ref{kappa_XI}. Recall that the value of imitation rate used in our simulations, unless otherwise stated, is $\kappa =  0.001$ as reported in Table 1. Here we presents results where this value of imitation rate is compared with a much smaller value of $\kappa = 0.00025$. In populations where the individuals imitate more quickly (i.e., higher $\kappa$), the oscillations in prevalence and vaccination dynamics arrive quicker with larger amplitude, although the converged vaccine uptake and disease prevalence are similar for higher and lower $\kappa$. For a higher $\kappa$, the convergence to  equilibria is quicker. These observations  are consistent with results reported by earlier studies \cite{bauch_2005_imi}. A smaller value of $\kappa$ ($\kappa=0.00025$) also altered behaviours of those individuals who are insensitive to prevalence change. Instead of converging to the pure non-vaccination equilibrium, the dynamics converge to a mixed endemic state, meaning that when imitation rate is very low, some individuals would still choose to vaccinate even when they are not very sensitive to disease prevalence. 

\begin{figure*}[htb!]
	\centering
	\graphicspath{{./fig/}}
	\includegraphics[width=\textwidth,trim={1.5cm 8cm 2.2cm, 3cm},clip]{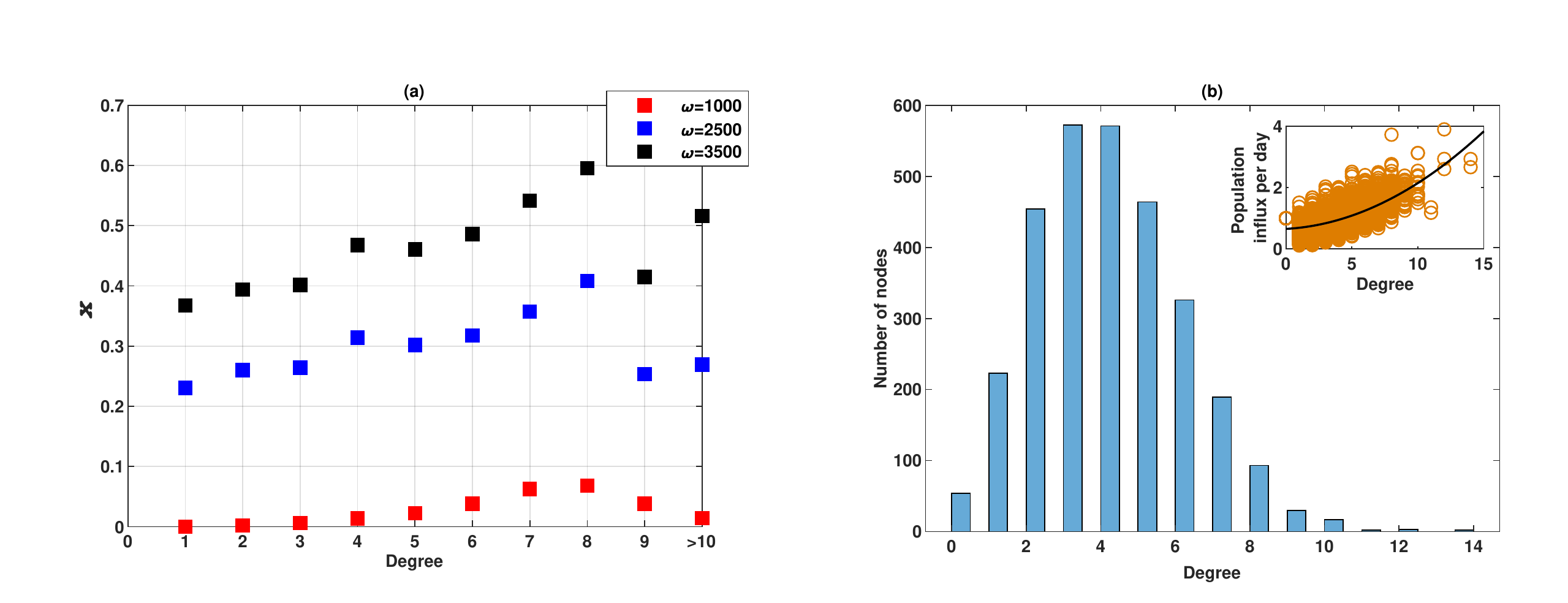}	
	\caption{The relationship between node degree and proportion of people who vaccinate voluntarily (vaccinating newborns only) for an Erd\"os-R\'enyi random network of 3000 nodes. (a) The fraction of vaccinated individuals as a function of the number of neighbouring suburbs each node has for three values of $\omega$. (b) Node degree in an Erd\"os-R\'enyi random network is distributed according to a Poisson distribution. The insert panel shows the population influx per node (as a sum of fractions from each source node) as a function of the node degree. }	
	\label{Vac_Nei}
\end{figure*}

For such a large network, vaccinating behaviour of individuals also depends on the weighted degree (number and weight of connections) of the node (suburb) in which they live. This is illustrated in Figure \ref{Vac_Nei}. For individuals living in highly connected nodes, there are many commuting destinations, allowing access to a broad spectrum of information on the local disease prevalence. Therefore, it is not surprising that we found  that individuals living in `hubs'  are more likely to get vaccinated, particularly when the population is sensitive to prevalence ($\omega$ is high), as shown in Figure \ref{Vac_Nei} (a). This observation is in accordance  with previous studies \cite{sto_network_2010}. Note that in  in Figure \ref{Vac_Nei} (a),  nodes with degree $k \geq 10$ are grouped into one bin to represent hubs, as the frequency count of these nodes is extremely low. Note also that there is a positive correlation between the number of degrees and the volume of population influx as measured by a sum of proportions from the source nodes (Figure \ref{Vac_Nei} (b)), indicating that a highly connected suburb has greater population influx on a daily basis.

\subsubsection{Vaccinating entire susceptible class}
\begin{figure*}[htb!]
	\centering
	\graphicspath{{./fig/}}
	\includegraphics[width=\textwidth,trim={3cm 0.5cm 3cm 0.5cm},clip]{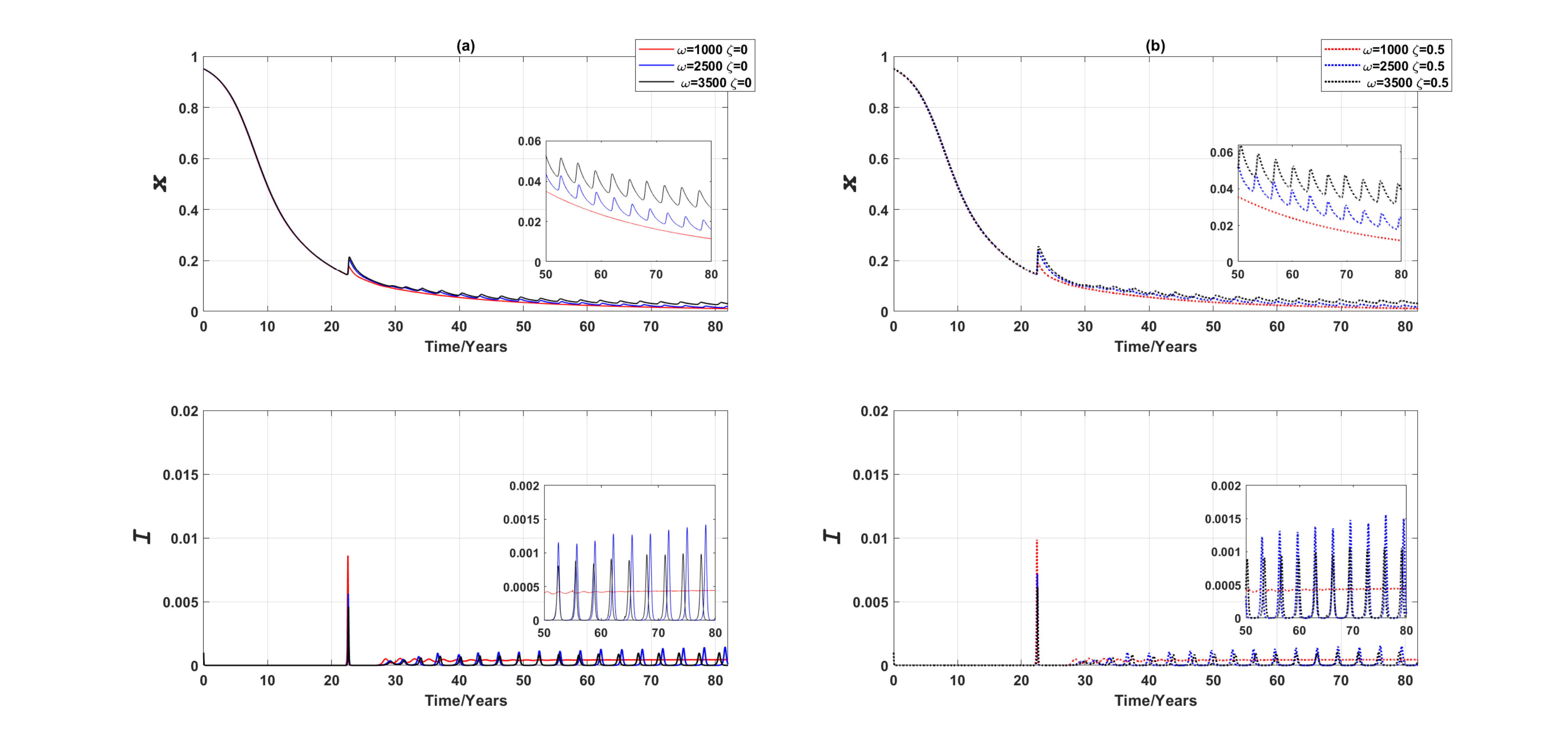}	
	\caption{Epidemic dynamics of a Erd\"os-R\'enyi random network of 3000 nodes  for three values of $\omega$ (vaccinating susceptible class regardless of age). The proportion of  vaccinated individuals, $x$, and the proportion of infected individuals, $I$, are shown against time. (a)$\zeta=0$ (b) $\zeta=0.5$ The inserted panel in each figure is a magnified section to show small oscillations. }	
	\label{3000_sus_XI}
\end{figure*}

If (voluntary) vaccination is offered to the  susceptible class regardless of age, we found that the global epidemic dynamics converge quicker compared to the similar scenario in the 3-node case. Only one predominant infection peak is observed, corresponding to the high infection prevalence around year 25, as shown in figure \ref{3000_sus_XI}.  Oscillations of small magnitude are observed for both vaccine uptake and disease prevalence  at later time-steps for middle or high values of $\omega$ (as shown in insets of figure \ref{3000_sus_XI}). These findings  also largely hold if half of the vaccines administered are ineffective (i.e., $\zeta=0.5$), although the predominant prevalence peak and the corresponding vaccination peak around year 25 are both higher than their counterparts observed for the case where $\zeta=0$.

\begin{figure*}[htb!]
	\centering
	\graphicspath{{./fig/}}
	\includegraphics[width=\textwidth,keepaspectratio,trim={4cm 1cm 4cm 1cm},clip]{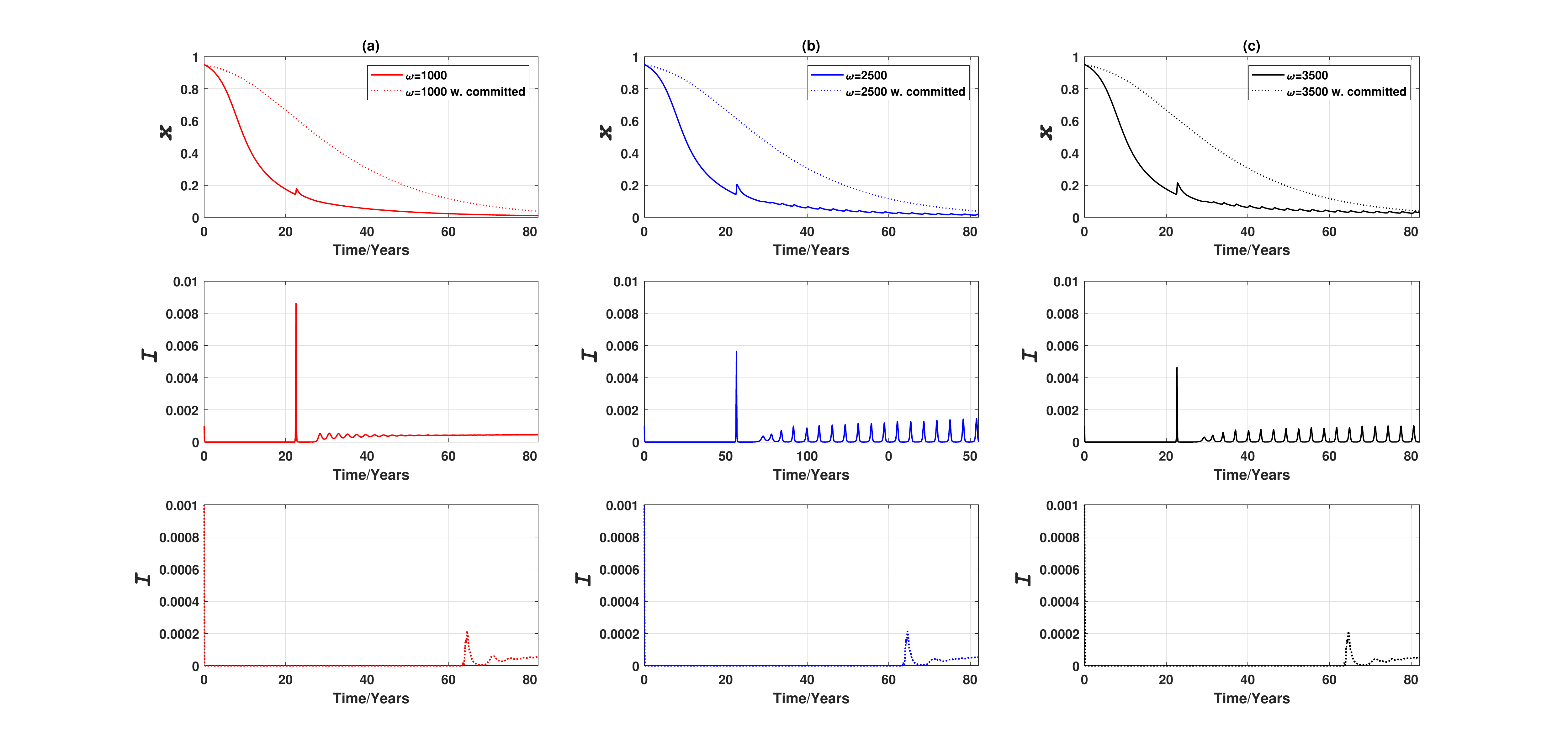}	
	\caption{Epidemic dynamics of a Erd\"os-R\'enyi random network of 3000 nodes  with committed vaccine recipients for three values of $\omega$ (vaccinating the entire susceptible class). (a) frequency of vaccinated individuals, $x$, against time (b)  proportion of infected individuals, $I$, against time. $x^c=0.0002$. The existence of committed vaccine recipients delays the predominant peaks and reduces magnitude of oscillation in the proportion of vaccine recipients in later stages.}	
	\label{committed0p05_XI}
\end{figure*}

We also found that employing a small fraction of committed vaccine recipients prevents major epidemic by curbing disease prevalence. Such a finding holds for all $\omega$ (i.e., regardless population's responsiveness towards disease prevalence) as the magnitude of prevalence is too small to make a non-vaccinated individual switch to the vaccination strategy (Figure \ref{committed0p05_XI}).

\section{Discussion and conclusion}
\label{conclusion}

We presented a series of SIR-network models with imitation dynamics, aiming to model scenarios where individuals commute between their residence and work across a network (e.g., where each node represents a suburb). These models are able to capture diverse  travelling patterns (i.e., reflecting local  connectivity of suburbs), and different vaccinating behaviours affecting the global vaccination uptake and epidemic dynamics. We also analytically derived expressions for the reproductive number $R_0$ for the considered SIR-network models, and demonstrated how epidemics may evolve over time in these models. 

We showed that the stable oscillations in the vaccinating dynamics are only likely to occur either when there is no population mobility across nodes, or only with limited commuting destinations. We observed that, compared to the case where vaccination is only provided to newborns, if vaccination is provided to the entire susceptible class,  higher disease prevalence and more volatile oscillations in vaccination uptake are observed (particularly in  populations which are relatively responsive to the changes in disease prevalence). A more expansive travelling pattern simulated in a larger network encourages the attractor dynamics and the convergence to the endemic, mixed equilibria, again if individuals are sufficiently responsive towards the changes in the disease prevalence. If individuals are insensitive to the prevalence, they are hardly affected by different vaccinating models and remain as unvaccinated individuals, although the existence of committed vaccine recipients noticeably delays the convergence to the non-vaccinating equilibrium. The presented models highlight the important role of committed vaccine recipients in actively reducing $R_0$ and disease prevalence, strongly contributing to eradicating an epidemic spread. Similarly conclusions have been reached previously  \cite{Bai2015}, and our results extend these to imitation dynamics in SIR-network models.

Previous studies drew an important conclusion that highly connected hubs play a key role in containing infections as they are more likely to get vaccinated due to the higher risk of infection in social networks \cite{immunization,sto_network_2010}. Our results complement this finding by showing that a higher fraction of individuals who reside in highly connected suburbs choose to vaccinate compared to those living in relatively less connected suburbs. These hubs, often recognized as business districts, also have significantly higher population influx as the destination for many commuters from other suburbs.  Therefore, it is important for policy makers to leverage these job hubs in promoting vaccination campaigns and public health programs.

Overall, our results demonstrate that, in order to encourage vaccination behaviour and shorten the course of epidemic, policy makers of vaccination campaigns need to carefully orchestrate the following three factors: ensuring a number of committed vaccine recipients in each suburb, utilizing the vaccinating tendency of well-connected suburbs, and increasing individual awareness towards the prevalence change.

There are several avenues to extend this work further. This work assumes that the individuals from different  nodes (suburbs) only differ in their travelling patterns, using the same epidemic and behaviour parameters for individuals from all nodes. 
Also, the same $\omega$ and $\kappa$ are used for all nodes, by assuming that individuals living in all nodes are equally responsive towards disease prevalence and imitation.
Realistically, greater heterogeneity can be implemented by establishing context-specific $R_0$ and imitating parameters for factors such as the local population density, community size \cite{measle_R0}, and suburbs' level of connectivity. For example, residents living in highly connected suburbs may be more alert to changes in disease prevalence, and adopt imitation behaviours more quickly. Different network topologies can also be used, particularly scale-free networks \cite{scale_free} where a small number of nodes have a large number of links each. These highly connected nodes are a better representation of suburbs with extremely high population influx (e.g., central business districts and job hubs). 
It may also be instructive to translate the risk perception of vaccination and infection into tangible measures to demonstrate the aggregate social cost of an epidemic breakout, and help policy makers to visualize the cost effectiveness of different vaccinating strategies and estimate the financial burden for public health care. This study can also be extended by calibrating the network to scale-free networks \cite{Alessio1} while incorporating information theory \cite{shannon}. Networks from real-life connectivity and demographic are also of strong interest to mimic a breakout in a targeted region \cite{AceMod,Zachresoneaau5294}, so that a model of vaccination campaign can demonstrate a reduction in the severity of epidemic. These considerations could advance this study to more accurately reflect contagion dynamics in urban environment, and provide further insights to public health planning.

\section{Appendix}
\subsection{Next Generation Operator Approach}
\label{app1}
The proposed model (\ref{system2.1}) can be seen as a finely categorized SIR deterministic model, so that each health compartment (Susceptible, Infected and Recovered) has $M$ sub-classes. Let us denote $\mathbf{I}=\{I_1, I_2,...,I_M\}$ to represent $M$ infected host compartments, and $\mathbf{y}$ to represent $2M$ other host compartments consisting of susceptible compartments $\mathbf{y_s}=\{y_{s1}, y_{s2},...,y_{sM}\}$ and recovered compartments $\mathbf{y_r}=\{y_{r1}, y_{r2},...,y_{rM}\}$.
\begin{equation}
\begin{aligned}
\frac{dI_i}{dt} & =\mathcal{F}_i(I,y_s)-\mathcal{V}_i(I,y_r) \\
\frac{dy_s}{dt} & =\mathcal{G}_{sj}(I,y_s) \\
\frac{dy_r}{dt} & =\mathcal{G}_{rj}(I,y_r) \\
\end{aligned}
\end{equation}
where $i \in (1,M)$ and $j \in (1,M)$. 

$\mathcal{F}_i$ is the rate at which new infection enters infected compartments and $\mathcal{V}_i$ is the transfer of individuals out of or into the infected compartments.

When close to disease-free equilibrium where $S=S^*=1$, the model can be linearized to:
\begin{equation}
\frac{dI}{dt}=(\mathbf{F}-\mathbf{V})I
\end{equation}
where $\mathbf{F}_{ij}=\frac{\partial \mathcal{F}_i}{\partial \mathcal{I_j}} (0,S^*)$ and $\mathbf{V}_{ij}=\frac{\partial \mathcal{V}_i}{\partial \mathcal{I_j}} (0,S^*)$\\

The next generation matrix, $\mathbf{K}$, is then given by:
\begin{equation}
\mathbf{K}=\mathbf{FV}^{-1}
\end{equation}
where each entry $K_{ij}$ represents the expected number of secondary cases which an infected individual imposes on the rest of the compartments. $\mathbf{F}$ and $\mathbf{V}$ are given as follows:
\begin{equation}
\mathbf{F}=
\begin{bmatrix}\frac{\partial \mathcal{F}_1}{\partial I_1} & \frac{\partial \mathcal{F}_1}{\partial I_2} & \cdots & \frac{\partial \mathcal{F}_1}{\partial I_M} \\
\frac{\partial \mathcal{F}_2}{\partial I_1} & \frac{\partial \mathcal{F}_2}{\partial I_2} & \cdots & \frac{\partial \mathcal{F}_2}{\partial I_M} \\
\vdots                            & \vdots                            & \ddots & \vdots \\
\frac{\partial \mathcal{F}_M}{\partial I_1} & \frac{\partial \mathcal{F}_M}{\partial I_2} & \cdots & \frac{\partial \mathcal{F}_M}{\partial I_M} \\
\end{bmatrix}
\end{equation}	

\begin{equation}
\mathbf{V}=
\begin{bmatrix}\frac{\partial \mathcal{V}_1}{\partial I_1} & \frac{\partial \mathcal{V}_1}{\partial I_2} & \cdots & \frac{\partial \mathcal{V}_1}{\partial I_M} \\
\frac{\partial \mathcal{V}_2}{\partial I_1} & \frac{\partial \mathcal{V}_2}{\partial I_2} & \cdots & \frac{\partial \mathcal{V}_2}{\partial I_M} \\
\vdots                            & \vdots                            & \ddots & \vdots \\
\frac{\partial \mathcal{V}_M}{\partial I_1} & \frac{\partial \mathcal{V}_M}{\partial I_2} & \cdots & \frac{\partial \mathcal{V}_M}{\partial I_M} \\
\end{bmatrix}
\end{equation}
The basic reproduction number $R_0$ is given by the most dominant eigenvalue (or `spatial radius' $\rho$) of $\mathbf{K}$ \cite{NGM1990,NGM2010,R0_beta}, and therefore:
\begin{equation}
R_0=\rho(\mathbf{K})
\end{equation}

\subsection*{Declaration of interest}
The authors declare that they have no competing interests.

\end{document}